\documentclass[11pt, a4paper,onecolumn]{article}

\usepackage{rotating}
\usepackage[table,xcdraw]{xcolor}
\usepackage{graphicx}
\usepackage{geometry}
\usepackage{lscape}
\usepackage{amsmath}
\usepackage{titlesec}
\titlespacing*{\section}
{0pt}{0.4ex plus 1ex minus .2ex}{0.4ex plus .2ex}

\usepackage[default]{lato}
\usepackage[T1]{fontenc}

\usepackage{xcolor}
\usepackage{amssymb}
\usepackage[font=footnotesize,labelfont=bf]{caption}
\usepackage{graphicx}
\usepackage{natbib}
\usepackage{hyperref}
\usepackage{url}

\title{\LARGE \bf The role of relatedness and strategic linkages between domestic and MNE sectors in regional branching and resilience}

 \author{Mattie Landman$^{1}$, Sanna Ojanperä$^{2,3}$, Stephen Kinsella$^{4}$, Neave O'Clery$^{5,3,1,\ast}$ \\
 \\ 1. Mathematical Institute, University of Oxford, UK \\
 2. Oxford Internet Institute, University of Oxford, UK \\
 3. The Alan Turing Institute, UK \\
 4. Kemmy Business School, University of Limerick, Ireland \\
 5. Centre for Advanced Spatial Analysis, University College London, UK \\
  }

\begin{document}

\date{\today}
\maketitle
\thispagestyle{plain}
\pagestyle{plain}

\vspace{0.25cm}
\noindent\rule{\linewidth}{0.35pt}
\vspace{0.25cm}

\setlength{\leftskip}{1.5cm}

A B S T R A C T 

Despite the key role of multinational enterprises (MNEs) in both international markets and domestic economies, there is no consensus on their impact on their host economy. In particular, do MNEs stimulate new domestic firms through knowledge spillovers? Here, we look at the impact of MNEs on the entry and exit of domestic industries in Irish regions before, during, and after the 2008 Financial Crisis. Specifically, we are interested in whether the presence of MNEs in a region results in knowledge spillovers and the creation of new domestic industries in related sectors. To quantify how related an industry is to a region's industry basket we propose two cohesion measures, \emph{weighted closeness} and \emph{strategic closeness}, which capture direct linkages and the complex connectivity structure between industries in a region respectively. We use a dataset of government-supported firms in Ireland (covering 90\% of manufacturing and exporting) between 2006-2019. We find that domestic industries are both more likely to enter and less likely to leave a region if they are related to so-called ‘overlapping’ industries containing both domestic and MNE firms. In contrast, we find a negative impact on domestic entry and survival from cohesion to 'exclusive MNE' industries, suggesting that domestic firms are unable to 'leap' and thrive in MNE-proximate industries likely due to a technology or know-how gap. This dynamic was broken, with domestic firms entering MNE exclusive sectors, by a large injection of Brexit diversification funds in 2017-18. Finally, the type of cohesion matters. For example, strategic rather than weighted closeness to exclusive domestic sectors matters for both entries and exits. 

\setlength{\leftskip}{0cm}

\vspace{0.3cm}
\noindent\rule{\linewidth}{0.35pt}
\vspace{0.3cm}

\setlength{\leftskip}{1.5cm}


\vspace{0.1cm}

\setlength{\leftskip}{0pt}

\vspace{0.3cm}
\noindent\rule{\linewidth}{0.35pt}
\vspace{0.3cm}
\clearpage

\onecolumn

\section{Introduction}

The avenues through which regions can generate economic prosperity and growth has long occupied a central position in the global research agenda. Foundational theories emerging from evolutionary economic geography suggest that regions grow by combining existing capabilities to create new economic activity \citep{nelson1982}. As it is costly to develop new activities that require capabilities that are unavailable in a region, regions tend to diversify into economic activities that are related to their current capabilities in a path dependent manner \citep{frenken2007theoretical, hidalgo2007product}. They may also look outwards to access new capabilities through external actors such as suppliers in neighbouring regions, migrants or foreign direct investment. In particular, attracting MNEs is seen as a key channel to \emph{import} new capabilities and generate knowledge spillovers via technology \citep{markusen1999foreign, arnold2009gifted} and skill transfer \citep{gorg2005spillovers, balsvik2011labor}. These spillovers are thought to enrich a region's capability base and thereby enhance domestic diversification opportunities. 

MNEs are generally viewed as beneficial to a host economy as they transfer financial resources \citep{iammarino_multinationals_2013}, create new market opportunities \citep{crescenzi_foreign_2015} and influence the productivity and innovation of co-located domestic firms through spillover effects \citep{iammarino_multinationals_2013}. Spillover effects can emerge through a variety of channels including demonstration effects, competition effects and labour mobility \citep{blomstrom1998multinational}. These are most often captured empirically through supply-chain linkages, but we focus on inter-sectoral labour mobility here in order to better proxy for knowledge spillovers \citep{balsvik2011labor, gorg2005spillovers}. These spillover effects may not always materialise, however, as MNEs actively protect their know-how and skills to prevent competition \citep{alcacer2016spatial}, or the capability gap may be too large between domestic and MNE firms and workers limiting absorptive capacity \citep{kokko1996local, blomstrom1998multinational}.    

In this study, we are interested in whether cohesion to MNEs in related sectors leads to knowledge spillovers that drive new domestic industry entries at a regional level. While a huge number of studies have investigated the effect of MNEs \textit{within} an industry \citep{gorg2001multinational, harris2003foreign, crescenzi_foreign_2015}, there have been fewer studies focusing on \textit{inter-industry} impacts. These include the impact of supply-chain linkages to MNEs on domestic entry \citep{ayyagari_does_2010}, and the cohesion of MNE entries to the local knowledge base \citep{elekes_foreign_owned_2019}. Here we focus specifically on the role of inter-industry knowledge spillovers from local MNEs for domestic industry entry.   

Further, we are interested in whether MNEs have a protective effect on regional domestic industry survival. Within the evolutionary economic geography literature, resilience is studied from an evolutionary perspective in which it is defined as a region's ability to successfully diversify into new growth paths when faced with an economic shock \citep{simmie_economic_2010}. The current consensus is that the more variety and the more closely an industry is related to a region's industrial basket, the more likely it is to survive \citep{nelson1982, neffke2011regions, balland_technological_2015}. In terms of inter-industry spillover effects from MNE to domestic industries at a regional level, 
\citet{szakalne2019agglomeration} found that the greater the variety of MNEs within a region's industrial portfolio the higher the chance of firm survival and the greater the region's resilience. Here we extend this literature, focusing on the impact of cohesion to MNEs in terms of knowledge or skill linkages on domestic industry exits. 

In order to quantify the potential for knowledge spillovers, encompassing a range of potential mechanisms, we deploy the skill-relatedness metric developed by \citet{neffke2013skill}. Skill-relatedness is a pair-wise measure of industry skill-similarity based on inter-industry labour mobility, and is used to build a cohesion measure which quantifies the degree of relatedness between an industry and the wider existing industrial basket of a region. Examples of such measures include the closeness measure \citep{neffke2011regions} and the product space density \citep{hidalgo2007product}, which have been widely used to study regional industrial diversification and structural change \citep{neffke2011regions, o2018skill, boschma2013emergence}. These measures, however, fail to consider the connectivity of related industries both between each other and with other industries in the region. In other words, they do not consider the high-order linkages that form a densely connected group of related industries in a region. Here we introduce a new measure, \textit{strategic closeness}, which captures cohesion to industries which are themselves well-connected thus quantifying relatedness \textit{between} sectors in a region.

We carry out our analysis on a subset of government-supported Irish firms that covers the vast majority of manufacturing and exporting firms in Ireland. Knowledge spillovers from MNEs are most likely to occur amongst these firms as they represent the most productive and complex part of an economy \citep{kokko_technology_1994}. Unlike previous studies, we separately investigate the impact of so-called 'overlapping industries' - those that have both MNE and domestic employment in a region - and MNE-only industries ('exclusive MNE industries') on domestic industry entry and exit. We focus on three distinct periods, before the Financial Crisis (2006-9), the recession (20010-14) and the recovery period (2015-19) which coincided with the Brexit referendum of 2016.  

We find that cohesion to overlapping industries is positively associated with both domestic industry entry and survival. In contrast, we find that if an industry is proximate to MNE-only industries it reduces the industry's chance of entry and survival. Our results suggest that domestic industries are unable to benefit from spillovers in this case due to a large technological and know-how gap. Hence, while it is difficult for domestic firms to 'leap' into these more complex and cognitively distant MNE dominated industries, once they have successfully entered and coexist with MNEs, overlapping industries appear to successfully induce further new entries via knowledge spillovers. We also find evidence that with a large amount of financial support, such as that injected in 2017-18 to enhance firm diversification in response to Brexit fears, domestic industries manage to make larger cognitive `leaps' and break into MNE dominated industries. Finally, we find that the type of cohesion matters. Being more deeply embedded with exclusive domestic industries, and thereby close to a denser concentration of inter-connected domestic industries, is associated with both more entries and better survival.  

We briefly provide an overview of the structure of the rest of the paper. Following a comprehensive literature review, we introduce the data and definitions of industry entries and exits as well as present some preliminary statistics and trends in the data. We then focus on the methodological development of the closeness measure of \citet{neffke2011regions} and the introduction of a new cohesion measure, the strategic closeness, before adapting both measures to account for domestic and MNE industries separately. We then present our econometric model, and our results. Finally, the paper concludes by discussing some potential policy implications of our work. 

\section{Literature}

\subsection{MNEs as agents of structural change}
Regional diversification is often depicted as a branching process in which a region develops new economic activities by drawing on and recombining capabilities, particularly know-how and skills embedded in workers, that are present within the region \citep{hidalgo2007product, frenken2007theoretical}. This is because search costs rapidly rise as the gap between the regionally available skills and know-how and those that are required for the new economic activity widens. Furthermore, new activities unrelated to the existing knowledge base of a region tend to have a lower probability of survival \citep{nelson1982, neffke2011regions}. Hence, related diversification (diversification into industries that are cognitively similar within a region) is the dominant channel for industrial diversification, while unrelated diversification (diversification into industries that are cognitively dissimilar) is rare \citep{frenken2007related, pinheiro2018shooting}. What is less clear within this literature is the role that external actors (e.g. suppliers and customers in neighbouring regions or foreign direct investment) play in the development paths of regions.

As the number and importance of MNEs has risen globally, many governments have developed industrial policies aimed at attracting FDI and other kinds of MNE engagement. This is as MNEs are seen as key generators of income, innovation and growth for the host economy. Examples of potential mechanisms through which MNEs bestow a beneficial effect on host countries' economies include directly via financial resources (spending on local suppliers, capital investment, employment, tax revenue), technology (R\&D), know-how in terms of management and training the workforce, as well as through linkages to value chains \citep{iammarino_multinationals_2013}. Furthermore, market access spillovers from MNEs to domestic firms are also important as they connect regions to global markets (and thereby induce domestic exporting activity) \citep{crescenzi2014innovation}. Our focus here, knowledge spillovers from MNEs to domestic firms have been suggested to primarily occur along three channels: demonstration effects where domestic firms gain knowledge by imitating MNE firms, competition effects, and knowledge transfer through labour mobility \citep{blomstrom1998multinational}. This dimension is particularly important with respect to regional industrial dynamics, and the import of know-how into a local labour force. 

Spillover effects to the domestic host economy may not always materialize and may even be negative \citep{gorg2004much, crespo2007determinant}. This can be attributed to MNE characteristics, which include actively protecting their know-how to reduce knowledge leakages to domestic competitors \citep{alcacer2016spatial}, or out-competing domestic firms on the labour market by providing better employment conditions to workers \citep{aitken1996wages, barry2005foreign}. A related branch of literature has specifically investigated how the `absorptive capacity' of domestic firms influences spillover effects \citep{kokko1996local, blomstrom1998multinational}. The absorptive capacity of a firm is defined as a firm’s ability to recognize valuable new knowledge, integrate it into the firm and use it productively \citep{zahra2002absorptive}. Various authors have argued that the lack of spillovers from MNEs to domestic firms is due to a wide skill or technology gap between the two groups \citep{kokko_technology_1994, girma2002there}. Empirical studies have shown that the strength of MNE-domestic spillover effects rise as the size and productivity of domestic firms increases \citep{bekes_spillovers_2009}. 

There have been a variety of empirical studies investigating how MNEs influence their host economy. These studies vary in the country they study, which aspect of economic development they consider (e.g., employment growth, industrial diversification, or firm productivity and innovation), and how they define MNE presence (FDI, value added to GDP, or measures related to R\&D expenditure, sectoral output, foreign equity, sales, employment etc.). As a result of diverging research findings, perhaps owing to the heterogeneous research designs employed, there is currently no consensus within the literature on the impact of MNEs on a host region. We list some of these studies in Table~\ref{Tab:Lit}. 

We focus here specifically on inter-industry spillovers. Within this burgeoning literature, we highlight a few studies of particular relevance which focus on domestic industry entry and cohesion to MNEs as proxied by inter-industry linkages. \citet{gorg_multinational_2002} and \citet{ayyagari_does_2010} find that the presence of related MNEs (supply-chain linkages) is positively associated with domestic entries in the manufacturing and service sectors respectively. \citet{lo_turco_local_2019} looked at product entries, finding that cohesion to MNEs enhances entry, particularly for more productive, established and local selling firms. Finally, \citet{smarzynska2004does, bekes_spillovers_2009} focus on firm productivity, similarly finding that cohesion to MNEs via supply chain linkages promotes productivity, particularly for larger firms and those focused on the domestic market.

We choose Ireland due to the country's profile as a highly developed open economy \citep{oleary2018} with substantial MNE presence and an FDI-oriented industrial policy. In fact, Ireland is one of the world's most active countries in terms of industrial policy. Key objectives include promoting export-led growth \citep{breathnach_regional_2015}, generating industry-university R\&D partnerships \citep{barry2014outward}, and developing better linkages between industries \citep{barry2014, oleary2018}. Some of the strategies developed to foster linkages and knowledge spillovers between domestic and MNE activities include the development of collaborative R\&D infrastructure and clusters \citep{department2014}, which have been linked to the emergence of industrial clusters \citep{oconnor_clustering_2017}. 

Various authors have investigated the role of MNEs on the Irish economy, however the impact of MNEs on domestic activities remains unclear. Studies have found that the presence of MNEs within the same sector influences the entry rate, productivity and employment growth of domestic firms \citep{gorg2005foreign}. Specifically, for productivity and employment growth, benefits have only been observed in high-tech domestic sectors \citep{gorg2003multinational}. In contrast, \citet{barry2003foreign} found a negative effect of MNEs on domestic exporting firms’ wages and productivity. Most related to our work, \citet{gorg2001multinational} suggested that related MNEs support domestic industry entrance through supply chain linkages, but \citet{di2018productivity} found a negative (or non-existent for R\&D-active firms) impact of MNE supply chain linkages on domestic firm productivity. 

Our study differs in five key aspects to previous work. First, we focus on cohesion between new domestic sectors and existing MNE industries as captured by inter-industry labour mobility patterns, a proxy for a broad range of potential knowledge spillovers. Secondly, we decompose MNE industries into two sets - overlapping industries (those with also domestic presence) and exclusive MNE industries. This enables us to distinguish the role of these distinct sets particularly with respect to their capability gap to domestic firms. Thirdly, we investigate which type of cohesion to MNEs matters. Specifically, as we outline below, we develop a new cohesion to capture the complex structure of linkages present between industries in a region. Fourthly, we conduct our analysis at a granular industry-region level, unlike the majority of studies which focus on industries at the national level, or regions neglecting the industry dimension. Finally, we investigate the role of MNEs in three different economic eras: before and after the 2008 financial crisis, and during a recent rapid domestic growth period fuelled by Irish government support linked to Brexit.

\begin{table}
\centering
\caption{Findings in literature investigating the impact of MNEs on industry entrance/exit, employment change, and knowledge spillovers}
\begin{tabular}{ |p{2.3cm}|p{5.7cm}|p{3cm}|p{2cm}|p{1.8cm}|  } 
\hline
Article & Impact of MNEs & Unit of analysis & Geography & Temporal coverage \\
\hline

\multicolumn{5}{|c|}{\cellcolor[HTML]{C0C0C0}Investigating industry entrance and exit} \\
\hline
\citet{elekes_foreign_owned_2019} & MNEs introduce more structural change than domestic firms & Manufacturing firms & Hungary, subregions & 2000-2009 \\
\hline
\citet{lo_turco_local_2019} & Impact of MNE knowledge conditional on insidership in the local market, product-specific knowledge and local firms’ absorptive capacity & Manufacturing firms & Turkey &  2005-2009 \\
\hline
\citet{zhu_host-country_2012} & Local density and experiential learning affect location choices of MNEs & Sample of Asian bank affiliates & United States & 1997-2003 \\
\hline
\citet{ayyagari_does_2010} & Positive horizontal and vertical entry spillovers in services, but not manufacturing, FDI country of origin matters & Sample of Czech firms & Czech Republic & 1994-2000 \\
\hline
\citet{bekes_spillovers_2009} & Productive firms benefit more, no larger impact on exporting firms & Manufacturing firms & Hungary & 1992-2003 \\
\hline
\citet{gorg_multinational_2002} & Positive effect of MNEs on entry of domestic firms & Manufacturing firms & Ireland & 1996-2002 \\
\hline

\multicolumn{5}{|c|}{\cellcolor[HTML]{C0C0C0}Investigating employment change} \\
\hline
\citet{cortinovis_multinational_2020} & Employment growth independent of input–output relations & employment across sectors & Europe, regions & 2008-2013 \\
\hline
\citet{boschma_related_2009} & Employment growth & employment, exports, and imports & Italy, provinces & 1995-2003 \\
\hline

\multicolumn{5}{|c|}{\cellcolor[HTML]{C0C0C0}Investigating knowledge and productivity spillovers} \\
\hline
\citet{csafordi2020productivity} & No impact on productivity & employer-employee linked labour flows & Hungary & 2003-2011 \\
\hline
\citet{di2018productivity} & Limited evidence for spillovers impacting productivity & manufacturing and service firms & Ireland & 2008-2014 \\
\hline
\citet{jordaan_producer_2017} & Positive economic spillovers, enhanced where technology gap & firms and their suppliers & Mexico & 2000-2001 \\
\hline
\citet{haller_domestic_2014} & MNEs impact productivity positively in transport, storage, and communication sectors, but negatively in wholesale and retail trade & all non-financial services firms & Ireland & 2001-2007 \\
\hline
\citet{barrios_spillovers_2011} & Spillovers have mixed effects for productivity depending on measure of backward linkages & manufacturing firms & Ireland & 1990-1998 \\
\hline
\citet{frenken2007related} & Spillovers enhance employment growth & employment except agriculture & Netherlands & 1996-2002 \\
\hline
\citet{ruane_foreign_2005} & Weak evidence of positive spillovers & manufacturing firms & Ireland & 1991-1998 \\
\hline
\citet{barry2005foreign} & Spillovers reduce wages and productivity for exporting firms, but have no effect on non-exporters & sample of manufacturing firms & Ireland & 1990-1998 \\
\hline
\end{tabular}
\label{Tab:Lit}
\end{table}

\subsection{Regional resilience}

Regional resilience has become a prominent focus area in the research and policy agenda, featuring in the target indicators of the Sustainable Development Goals. Indeed, Target 1.5 aims to \emph{‘By 2030 build the resilience of the poor and those in vulnerable situations, and reduce their exposure and vulnerability to climate-related extreme events and other economic, social and environmental shocks and disasters’} \citep{bahadur_resilience_2015}. A growing literature investigates the determinants of a region's ability to adapt to an external shock such as the Financial Crisis \citep{martin_are_2016, crescenzi_geography_2016, fratesi_crisis_2016, xiao_resilience_2018}. However, despite the growing popularity of resilience in the research and policy agenda, there are concerns over the usefulness of the concept stemming from lack of clarity on its definition \citep{martin_regional_2012}; the appropriate theoretical and empirical frameworks to measure and analyze it \citep{bristow_crisis_2015, diodato_resilience_2015, faggian_regional_2018}; its determinants \citep{martin_regional_2012}; and tools to design or implement appropriate policies. 

While the concept of resilience is multidimensional and has been interpreted in various ways, there are three main conceptual approaches: engineering, ecological, and adaptive \citep{simmie_economic_2010, martin2015notion}. Scholars advocating an \emph{engineering-based approach} emphasise the ability of an economic system to return to its stable or pre-crisis equilibrium state after a crisis. The \emph{ecological-based view} concerns the magnitude of a shock that a system can weather without shifting to a new equilibrium state. The \emph{evolutionary approach} departs from these equilibrium-based frameworks and defines resilience as the ability of an economy to successfully diversify and branch out into new growth paths, thereby countering economic decline \citep{martin_resilience_2019, boschma_towards_2015}. In this study we adopt the latter approach and study regional resilience through industry exits, defining the exit of an industry from a region via a drop in employment below a threshold \citep{neffke2011regions}. Employment-based measures are thought to reflect the societal impacts of a crisis more readily than output variables such as GDP or GVA \citep{diodato_resilience_2015}.

A range of studies have investigated the relationship between the industrial portfolio of a region and industry exits \citep{neffke2011regions, essletzbichler_relatedness_2015, szakalne2019agglomeration}. Studies have found that a region with a high variety of industries is better able to adapt to sector-specific shocks \citep{boschma_towards_2015,szakalne2019agglomeration}. Furthermore, \citet{balland_technological_2015} showed that regions with a high degree of relatedness to existing technologies in which the region does not have comparative advantage (competitive presence) had a greater capacity to weather technological crisis. Overall, there is evidence in the literature to support both industry variety and relatedness as key factors in industrial resilience. 

Here, we are primarily interested in whether cohesion to MNE industries enhances domestic industry survival and regional resilience. MNE-domestic linkages have a protective effect on domestic industries in times of crisis via a number of mechanisms. Similar to above, knowledge spillovers can occur through demonstration effects, labour mobility and competition channels. Primarily, knowledge and experience built up by MNEs on external shocks may transfer to domestic firms \citep{fainshmidt2017mne} via demonstration effects and labour mobility. Labour mobility may also assist via reallocation of workers between sectors in a region \citep{diodato_resilience_2015}. An alternative mechanism, productivity gained from technological and knowledge spillovers is expected to reduce the average cost of domestic production which in turn may help firms to survive in the face of economic shocks \citep{gorg2003multinational}. 

Very few empirical studies have investigated the role of MNEs in regional resilience. Closest to our work, \citet{szakalne2019agglomeration} showed that a greater variety of MNE industries reduces the likelihood of domestic firm exit, using data on Hungarian regions. This effect was particularly strong for regions undergoing economic transition. Focusing on Ireland, \citet{gorg2003multinational} found that the presence of MNEs in the same industry increases domestic firm chance of survival through technological spillovers. This was only significant for high-tech sectors, while no effect of MNE presence was found for low-tech sectors. We add to this literature by investigating if the cohesion to MNE industries, measured at an industry-region level, provides a protective effect. 

\subsection{Industrial Cohesion}

In this study, our aim is to investigate how the presence of MNE (and domestic) industries within a region's industrial basket impacts the entry (or exit) of an industry. We are therefore interested in measuring the capability or knowledge-distance between an industry and a region's current industry basket. Within the literature, cohesion is defined as the degree of relatedness amongst industries within a region and a measure of the opportunity for knowledge spillovers \citep{neffke2011regions, frenken2007related}. Cohesion measures are typically used to quantify the degree of structural change induced by an industry as it enters or leaves a region. When an industry enters (or exits) a region, it brings new (or removes current) capabilities. Hence, according to which industries enter or exit, and how they are related to the current industrial portfolio, they differently impact the region through changes in the combined total of the region's capabilities. 

Cohesion measures are typically derived from the structure of an industry network, also referred to as an `industry space' \citep{neffke2017, hausmann2007}. This is a network where nodes represent industries and edge-weights correspond to the degree of capability-overlap between industry pairs. The advantage of using an industry network is that it allows for the topological structure resulting from the relatedness amongst all industry pairs to be analysed via a complex systems approach. To construct an industry network, a measure of relatedness amongst industries is required. Since the true level of capability overlap cannot be directly measured, an outcomes-based approach is taken to infer the degree of relatedness. This type of approach varies according to the data source considered and capability type. For example, the co-location of industry pairs on patents has been used to measure the degree of technological-relatedness \citep{ellison2010causes, jaffe1989characterizing} and supply chain (IO) linkages have been used to estimate the degree of supplier-buyer sharing or similarity between industries \citep{acemoglu2015networks}. 

In this study we adopt the skill-relatedness index which is based on labour mobility between industries \citep{neffke2013skill} as we primarily focus on knowledge spillovers and labour pooling between MNEs and domestic firms. By using labour mobility to infer skill-relatedness we assume that if two industries share a high degree of skills and knowledge, workers will more freely move between them. This is because a worker's skill set from one of these industries will also be highly valued within the other industry and thereby be most likely to switch to this industry (compared to others). Various authors have argued that the skill-relatedness index is the best approach to model regional growth due to the key role of tacit know-how and skills embedded within workers \citep{o2021unravelling}.  

One of the first cohesion measures was introduced by \citet{neffke2011regions}. The authors defined the \textit{closeness} of an industry as the count of the number of related industries (neighbours in the industry network) present within a region's industrial portfolio. This effectively captures how connected or embedded an industry is to other industries in the region. The authors found that for manufacturing industries in Sweden, industries that enter a region have a higher closeness, while those that exit have a lower closeness. Another well-known cohesion measure is the \textit{density} (or related employment\footnote{Related employment is very similar to the density measure but also considers the employment size of each industry.}) measure. This metric measures the strength of relatedness (edge weight) between an industry and its neighbours relative to the strength of relatedness to all sectors. It has been used in various applications to predict regional industry diversification and employment growth \citep{neffke2011regions, neffke2013skill, o2018skill, boschma2013emergence, hausmann2007}. In all of these studies, industries that have a higher density are found to be more likely to enter a region.

Both of these variables are one-step measures in the sense that they only consider direct neighbours in the network. Consequently, all neighbours are treated homogeneously. What these measures fail to capture is the importance of the connectivity and embeddedness of their neighbours (the greater industry network structure). Various authors have argued that it is not only the presence of a related industry but also the assemblage of these industries, as well as other industries present in the region, that generate collective efficiency and further knowledge spillovers \citep{marshall_principles_1920}. Hence, spillovers are generated from the presence of a cluster of densely connected economic activity around an industry \citep{porter2011competitive}. Therefore, being more deeply embedded into the industry network enhances the chance of spillovers. A number of influential studies have also argued that economic activity that is more distant from an industry has shown to enhance innovation, however the economic activities cannot be too cognitively distant that learning cannot occur \citet{nooteboom1999innovation, frenken2007related}. In this study, we develop a new cohesion measure that captures both the presence of related industries as well as their connectivity to other industries within a region. The measure therefore captures the impact of higher-order linkages that may occur through the broader concentration and inter-connectivity of economic activity but is not too skill-distant from an industry.

Another cohesiveness measure, \textit{related variety}, has also been widely adopted within the literature \citet{frenken2007related, szakalne2019agglomeration}. A region with a high related variety has employment spread over a variety of industries within a few sectors, while one with low related variety has employment spread over industries within different sectors. The authors hypothesised that regions benefit from employment distributed in a variety of industries as more variety implies more potential for spillovers. However, the variety should primarily occur amongst industries in the same sector as limited spillovers occur amongst industries in different sectors. The measure is directly computed at a region level and is based on an entropy-calculation for employment distributed within industries spread across sectors. 

A central disadvantage of the related variety measure is that it relies on the hierarchical structure of the standard industrial classification system as a measure of relatedness. Thereby, when considering the related variety of a region, all industries within the same sector (2-digit industry class) are assumed to have the same relatedness. Secondly and most importantly for our study, the measure is calculated at a regional level making it less suited to our application. Our new metric, however, does capture some of the ideas behind related variety in that it identifies the presence of groups of related industries - although in our case we quantify relatedness via the skill-relatedness measure rather than the official industrial classification - in a region. The key difference, however, is that it enables us to look at the cohesion of a particular industry to these groups, resulting in an industry-region level variable. 

\section{Data and definitions}\label{Sec:Data}

\subsection{Industry data}

For this study, we use data covering the majority of exporting and manufacturing firms within the Irish economy. The data derives from the Irish Department of Business, Enterprise, and Innovation Annual Business Survey of Economic Impact, and includes firms assisted by the three Irish enterprise development agencies: Industrial Development Authority (IDA) Ireland, Enterprise Ireland, and Údarás na Gaeltachta. 

The dataset covers the period 2006-2019, and includes total employment (in assisted firms) at an industry-region-year level of aggregation. This is further broken down by firm ownership-type level (either Irish or foreign). Industries correspond to 4-digit NACE 2 industry level, and regions are at NUT III regional level. Further data descriptors across region, time and ownership type are presented in Table~\ref{Tab_DataDes} in Appendix~\ref{Appendix_Data}.

The dataset includes approximately 80\% of all manufacturing employment and 7\% of services employment within Ireland. Furthermore, it accounts for 90\% of total merchandise exports as well as 70\% of services exports (which comprise of approximately half of total Irish exports) \citep{breathnach_regional_2015}. It also includes approx. 63\% of the employment in all foreign-owned firms in Ireland. The dataset therefore covers both the majority of domestic and foreign manufacturing and exporting firms. As these firms are typically highly productive, complex and export-focused there is a higher likelihood of MNE-domestic knowledge spillovers occurring amongst them \citep{kokko_technology_1994, bekes_spillovers_2009}. These firms also act as leading drivers of economic development thus also offering an important indicator of regional economic prosperity.

\subsection{Industry presences, entries and exits}

We start by dividing our 14-year time-span into three time periods, namely 2006-2009, 2010-2014 and 2015-2019. Each of these periods can be associated with a distinct economic era in Ireland. The 2006-2009 period falls largely before the 2008 financial crisis started to take effect. A recession then characterised the 2010-2014 time period \citep{conefrey2018modelling}. Finally, the 2015-2019 period was characterised by fast growth of the domestic economy fuelled by Irish industrial policy support in response to Brexit \citep[Chapter 1]{roche2016austerity}. 
 
In this study, we investigate the entry and exit of domestic industries with respect to their cohesion to three mutually exclusive sets of existing industries within a region. These sets are `exclusive MNE' industries in which only MNE firms are active, 'exclusive domestic' industries in which only domestic firms are active and 'overlapping' industries in which both MNE and domestic industries are active. 
 
We define the presence, entry and exit of an industry $j$ in region $r$ at time $t$ as follows:
\begin{itemize}
    \item An MNE industry is present if there are more than $5$ employees in foreign-owned firms ($X_M(j,r,t)=1$ and otherwise 0), while a domestic industry is present if there are more than $5$ employees in Irish-owned firms ($X_D(j,r,t)=1$ and otherwise $0$).
    \item Similarly, an exclusive domestic industry is present if there are more than $5$ employees in Irish-owned firms and no employment in foreign-owned firm ($X_{excl D}(j,r,t) = 1$ and otherwise $0$). Similarly, an exclusive multinational industry is present if there are more than $5$ employees in foreign-owned firms and no employees in Irish-owned firms ($X_{excl M}(j,r,t)=1$ and otherwise $0$). For the case of overlapping industries, the industry is present if there are both more than $5$ employees present within Irish-owned firms and within foreign-owned firms ($X_{overlap}(j,r,t) = 1$ and otherwise $0$).
    \item A domestic industry entrant is an industry that had less than $5$ employees in the beginning of the time period, and then becomes present in the industrial portfolio of a region at the end of the time period  $(X_{D}(j,r,t)=0 \cap X_{D}(j,r,t+1)=1)$. 
    \item A domestic industry exit is an industry that had more than $5$ employees in the beginning of the time period and then was no longer present in the industrial portfolio of the region at the end of the time period $(X_{D}(j,r,t)=1 \cap X_{D}(j,r,t+1)=0)$.
\end{itemize}

We choose $5$ employees as our threshold measure to indicate the presence of an industry within a region \citep{neffke2018}, and hence less well established and potentially dormant industries are removed. 


\subsection{Skill-relatedness matrix}\label{Sec_Relatedness}

We use a second dataset to measure the skill-relatedness between industry pairs, following \citep{neffke2017}. \citet{o2019modular} previously constructed this relatedness measure for Irish industries using an anonymised administrative dataset from Ireland's Central Statistics Office. The dataset contains the employment records\footnote{The employment records are constructed from SPP35 annual tax returns filed by employers on their employees to the Irish Revenue Commissioners.} of each registered employee within the Irish formal economy. The dataset covers the 2005-2016 period\footnote{Note that there is an overlap in the time period to the above mentioned dataset. As the datasets differ and the skill-relatedness value has been shown to remain relatively constant across smaller time periods \cite{neffke2017}, we do not consider this to be a problem.}.

The skill relatedness matrix is constructed by following workers as jump between industries (4-digit NACE 1.1 industry codes). Entry i,j in this matrix corresponds to the average number of workers that transitioned between industry i and j per year between 2005-2016 normalised by the number that would have been expected to switch at random given the size of the industries. We convert the matrix to NACE 2 using the methodology of \citet{diodato2018network}. More details on the construction of the matrix and conversion steps can be found in the Appendix~\ref{SecSR} and Appendix~\ref{Appendix_Data} respectively. 

\begin{figure}[t!]
     \centering
     \includegraphics[width = \textwidth]{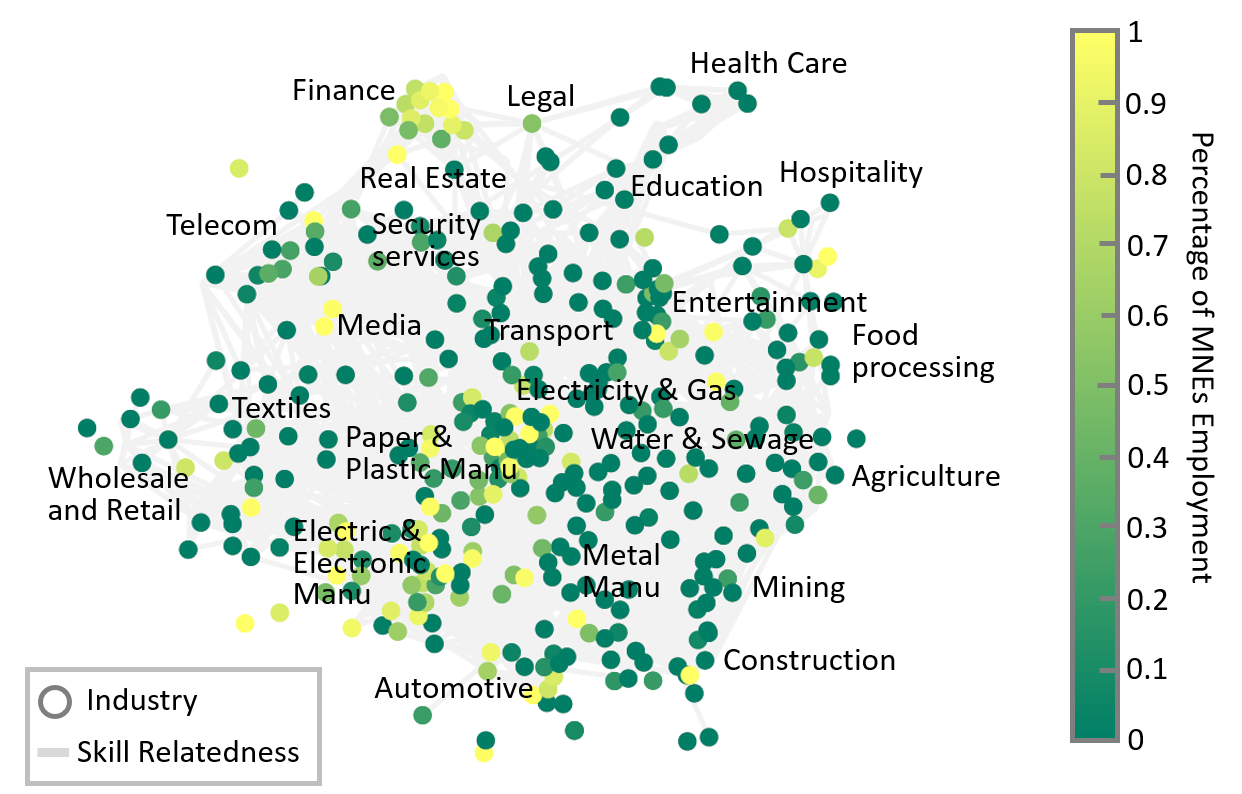}
     \caption{A visualization of percentage of MNE employment in each industry on the skill-relatedness network for Ireland. Each node represents an industry and each edge a skill relatedness linkage. Nodes are coloured according to the percentage of MNE employment within each industry. The network layout was generated using 'Force Atlas' in Gephi - a spring algorithm in which related industries are positioned closer together.}
     \label{Fig:SRN}
\end{figure}

The matrix $A_{SR}$ is the adjacency matrix of the skill-related industry network. We visualize the network shown in Figure~\ref{Fig:SRN}. Each node represents an industry and each edge a skill relatedness linkage (as encoded in $A_{SR}$). A spring algorithm called `Force Atlas' in Gephi is used to generate the spatial layout of nodes with more related industries positioned closer together. We have added a general labelling of groups of industries on the network for orientation, as well as coloured the nodes by the percentage of MNE employment. We observe that MNEs are concentrated mainly within the finance and high-tech manufacturing sectors. 

\section{Measuring the cohesiveness of an industry}

Here we introduce the \textit{weighted closeness} and \textit{strategic closeness} of an industry to the existing industrial basket of a region. 

\subsection{Weighted closeness}
The weighted closeness ($WC$) of an industry is similar to the closeness measure of \citet{neffke2018} in that it captures the number of related industries (to that industry) that are present in a region. This measure quantifies the cohesion of an industry to the industrial basket of a region as the number of related industries present in the region weighed by their relatedness.

Let the relatedness between industries be encoded in matrix $A_{SR}$\footnote{While we use the skill-relatedness measure of \citep{neffke2013skill} within our analysis, the cohesion measures can be used with any type of relatedness measure.}. the $WC$ measure is then given by, 
\begin{equation}\label{Eq_WC}
    WC(i,r,t) = \sum_{j,j\neq i} A_{SR}(i,j) \times X(j,r,t),
\end{equation}
where $X(j,r,t)=1$ if industry $j$ is present in region $r$ at time $t$ (and otherwise 0). 

Similar to closeness \citep{neffke2018}, weighted closeness only considers the presence of directly related industries, and does not take into account the wider 'global' structure of the industry network. Furthermore, although we weigh the presence of each directly related neighbour by its relatedness, the presence of each neighbour is treated homogeneously. Hence, the measure does not consider the connectivity of these neighbouring industries to other related industries within the network. Next, we introduce a new measure that is able to capture these higher-order connections. 

\subsection{Strategic closeness}

We propose a new measure, strategic closeness ($SC$), which does not only consider directly related industries (as in the case of the $WC$) but also their connectivity to both each other and other industries present within the region. In other words, the measure picks up the presence of higher order connections (of two steps away) in the local industrial basket. An industry with high $SC$ is not only related to industries in the region but these industries are themselves highly connected to both each other and other industries in the region. These are in a sense 'strategic' or highly embedded neighbours. These more distant industries increase the variety of skills and know-how an industry has access to, and are thought to promote innovation \citep{frenken2007related, nooteboom1999innovation}. 

The $SC$ measure models regional diversification as a diffusive process \citep{frenken2007theoretical}, and can be seen as a multi-step generalisation of the aforementioned $WC$ measure. Intuitively, it can best be understood by considering a random walker on the industry network. The random walker is initially positioned on the network with a uniform probability distribution across all industries present within the region's portfolio and is then allowed to move on the network. The walker jumps from one industry to another with probability proportional to the edge weight (relatedness) connecting them. The probability distribution for the 'location' of the walker (across all nodes) at a given 'time' (number of steps) captures the potential 'spreading' of the walker across the network in that number of steps, and is governed by the connectivity structure of the network. We enable the walker to move within the network for two steps, and hence the probabilities correspond to a cohesion measure that takes into account the presence of 'neighbours of neighbours' in the network. 

More formally, the industry network is defined via adjacency matrix $A_{SR}$ with entries corresponding to the relatedness between industry pairs. We define the degree vector $\mathbf{d}$, where $d(i)$ calculates the sum of all edge-weights that are connected to node $i$. This is given as $d(i) = \sum_j A_{SR}(i,j)$. The diagonal matrix of degrees is then defined as $D = \text{diag}(\mathbf{d})$. A random walker process on the industry network can now be defined as an associated Markov chain in which the probability of leaving a node is split amongst the edges of a node according to their relative weight. The transition probability for an edge connecting industry $i$ and $j$ is given by $A_{SR}(i,j)/d(i)$. Hence,
\begin{equation}
    \mathbf{p}_{\tau + 1} = \mathbf{p}_{\tau} D^{-1} A_{SR},
\end{equation}
where $\mathbf{p}_{\tau}$ $\in$ $\mathbb{R}^N$ a probability vector representing the probability of finding a random walker at node $i$ at time step $\tau$. Note that given an initial probability vector $\mathbf{p}_0$, the process can also be described as 
\begin{equation}
    \mathbf{p}_{\tau} = \mathbf{p}_{0} (D^{-1} A_{SR})^{\tau},
\end{equation}

Now, we define the starting probability of a random walker for region $r$ and at the base period ($t_{base}$) as the uniform distribution across all industries that are present in the region, this is given as: 
\begin{equation}
\mathbf{p}(i,r,t_{base}) = \frac{X(i,r,t_{base})}{\sum_j X(j,r,t_{base})}
\end{equation}
Finally, as we consider a two-step random walker process, we let $\tau = 2$ and define the strategic closeness vector as
\begin{equation}\label{Eq_SC}
    \mathbf{SC}(:,r,t)= \mathbf{p}(:,r,t_{base}) \, (D^{-1} A_{SR})^{2},
\end{equation}
where \: denotes all elements of the vector. The strategic closeness of industry $i$ is then defined as $SC(i,r,t)$. 

\begin{figure}[t!]
    \centering
    \includegraphics[width =0.8\textwidth]{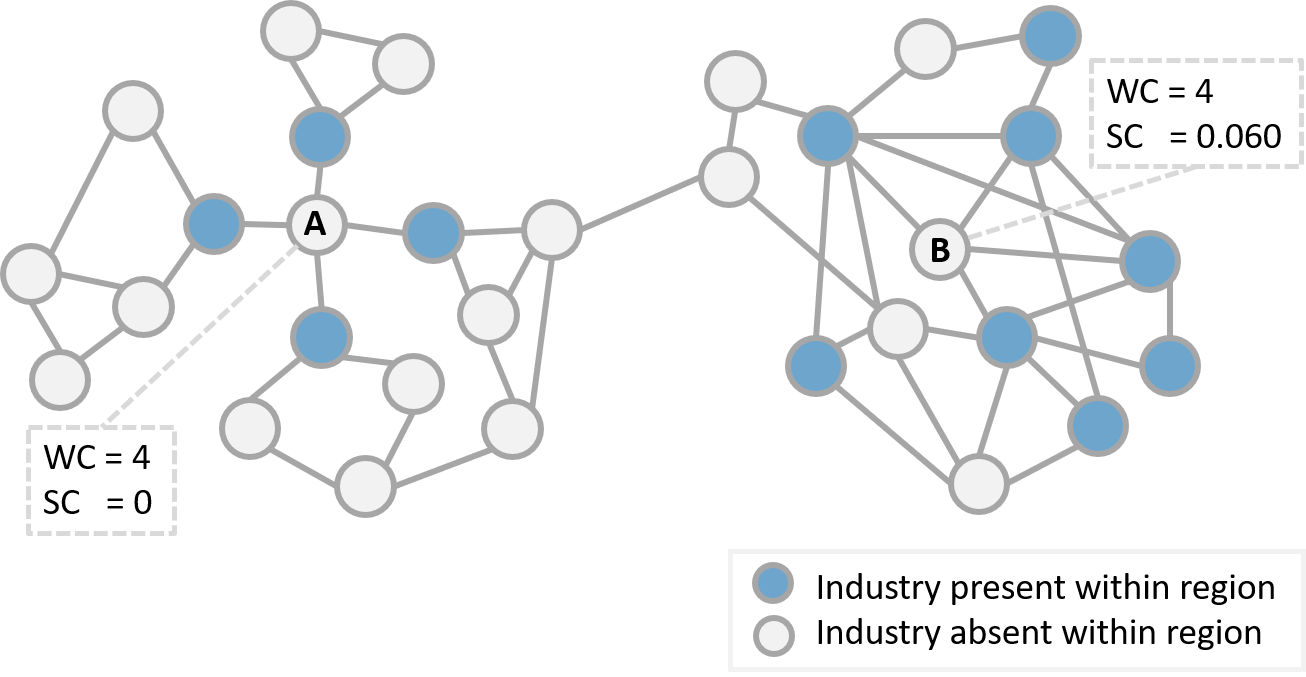}
    \caption{The comparison of two industries' (A and B) cohesion to the industrial portfolio of their region using the weighted closeness and strategic closeness metrics. The network represents a toy industrial network on which the mock region's industrial portfolio is shown. Each node represents an industry and each edge the level of relatedness between the corresponding two industries. Blue nodes are industries that are present within the region, while grey ones are absent.}
    \label{Fig:CohesionTogether2}
\end{figure}

We illustrate the complementary information gained from using the $SC$ measure alongside the $WC$ measure in Figure~\ref{Fig:CohesionTogether2}. Here we show the industrial portfolio of a mock region displayed on an industry network. For this example, an unweighted industry network is used. Nodes in blue represent industries that are present within the region, while those in grey are absent. We observe that both node $A$ and node $B$ are directly related to four other industries that are present within the region and hence both of these industries have the same $WC$. However, we can easily see that industry $B$ is directly connected to industries that are themselves both inter-connected and connected to other industries present in the region. Industry $B$ is therefore connected to a larger agglomeration of related industries which could provide access to a larger range of capabilities and opportunities to develop a variety of linkages. Hence, industry $B$ has a higher $SC$ cohesion value than industry $A$ (which has an $SC$ of zero as its related industries are not connected to each other or other industries present in the region). 

\subsection{Cohesion to domestic and MNE industries}\label{Sec:KeyMetrics}

Here we adapt $WC$ and $SC$ to account for the presence of exclusive domestic industries, exclusive MNE industries and overlapping industries separately.

First, we adapt the weighted closeness measure to include only exclusive domestic industries within the industry portfolio of a region. This is given by, 
\begin{equation} \label{Eq_WC_D}
    WC_{excl D}(i,r,t) = \sum_{j\in N_i,j\neq i} A_{SR}(i,j) \times X_{excl D}(j,r,t).
\end{equation}
where $X_{excl D}(j,r,t)=1$ if industry $j$ contains only domestic employment in region $r$ at time $t$. This measure captures the cohesiveness of industry $i$ to exclusive domestic industries in region $r$. The measure is analogously defined with respect to the presence of exclusive multinational industries (denoted as $WC_{excl M}(i,r,t)$) and the presence of overlapping industries (denoted as $WC_{overlapping}(i,r,t)$) within a region.

We similarly adapt the strategic closeness measure to capture the presence of different types of industries in a region. In the case of exclusive domestic industries, the starting probability of the random walker on industry $i$ within region $r$ at base time $t$ is defined as 
$$
p0_{excl D} (i,r,t) = \frac{X_{excl D}(i,r,t)}{\sum_j X_{excl D}(j,r,t)}
$$ 
and then 
\begin{equation}
  \mathbf{SC_{excl D}}(:,r,t) = \mathbf{p0_{excl D}}(:,r,t) \, (D^{-1} A_{SR})^{2}.
\end{equation}
where $\mathbf{p0_{excl D}}$, $D$ and $A_{SR}$ are defined as before. The strategic closeness of industry $i$ is then given as $SC_{excl D}(i,r,t)$. The measure is similarly defined for the strategic closeness to exclusive multinational industries (and denoted $SC_{excl M}(i,r,t)$) and for the strategic closeness to overlapping industries (denoted as $SC_{overlap}(i,r,t)$) within a region. 


\section{Econometric framework}\label{Sec:Model}

Our aim is to investigate the relationship between domestic industry entry (and exit) and the cohesiveness of the industry to exclusive MNE, exclusive domestic or to overlapping industries in the region across three distinct periods. To detect these relationships we set up a panel probit regression model in a similar frame to \citet{neffke2011regions} and \citet{szakalne2019agglomeration}.

For our first model, we investigate the relationship between domestic industry entrants and their cohesion to the different types of industries within the industry portfolio of the region. We run a fixed effects panel probit model for each of the three time-periods separately. The model is given by:
\begin{multline}
\label{probit1}
    \text{Probit}(\text{Pr}(\text{Entry}(i,r,t)=1|Z(i,r,t-1),\omega X_M (i,r,t-1),\gamma(i),\tau(r))) \\ = \phi(\alpha + \beta Z(i,r,t-1) + \omega X_M (i,r,t-1) + \gamma(i) + \tau(r) + \epsilon(i,r,t)),
\end{multline}
where $\phi(\cdot)$ is the cumulative distribution function of the standard normal distribution, and $Z(i,r,(t-1))$ is the value of the explanatory variable (cohesiveness measure) included in the model. This can be the cohesion to exclusive MNEs ($WC_{excl M}$, $SC_{excl M}$), exclusive domestic industries ($WC_{excl D}$, $SC_{excl D}$) or overlapping industries ($WC_{overlap}$, $SC_{overlapping}$). $\beta$ is then coefficient of the cohesion explanatory variable. $X_M$ indicates whether an MNE industry is already present, with corresponding coefficient vector $\omega$. $\gamma(i)$ and $\tau(r)$ are industry and region fixed effects, respectively. Through these fixed effects we control for within-region and within-industry variance. Regional fixed effects account for the number of MNE or domestic industries present within a region. We also include a coefficient term to absorb other dependencies that we have not controlled for. Furthermore, we include a standard robust error term. In this model we only consider domestic industries which are not yet present within a region as an observation. 

In our second model, we investigate the relationship between the exit of domestic industries and their cohesion to exclusive MNEs, exclusive domestic or to overlapping industries within a region. Using a very similar model as previously, we run a fixed effect panel probit model for the various time periods, given by:
\begin{multline}
\label{probit2}
    \text{Probit}(\text{Pr}(\text{Exit}(i,r,t)=1|Z(i,r,t-1),\omega X_M (i,r,t-1),\gamma(i),\tau(r))) \\ =\phi(\alpha + \beta Z(i,r,t-1) + \omega X_M (i,r,t-1) + \gamma(i) + \tau(r) + \epsilon(i,r,t)),
\end{multline}
where variables are similarly defined as in the first model. Here we only consider domestic industries that are already present within a region as an observation. 


\begin{figure}[t!]
    \centering
    \includegraphics[width = 0.7\textwidth]{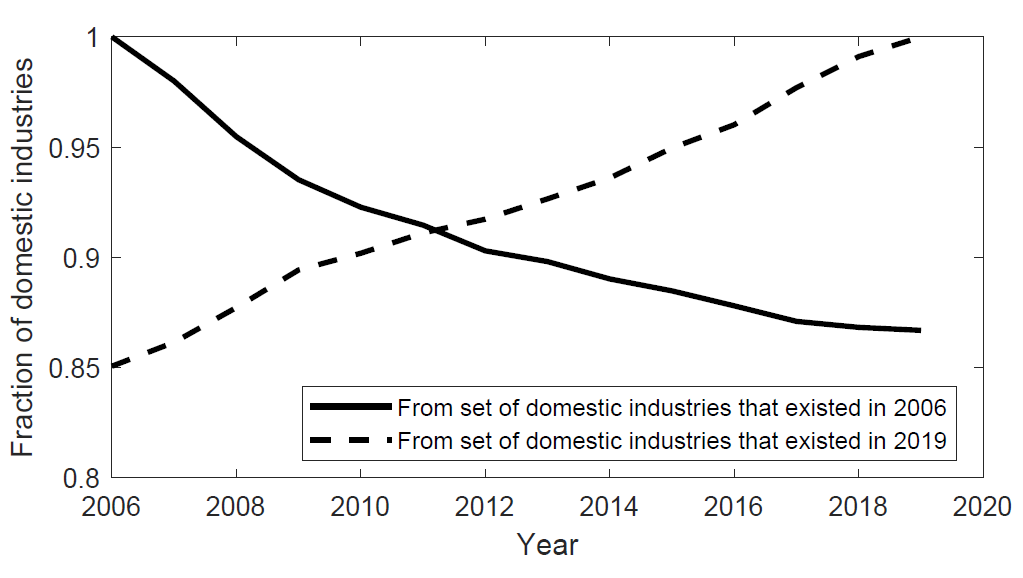}
    \caption{Graph showing the domestic structural change in Irish regions between 2006--2019. The solid line shows, for all regions in Ireland, the share of domestic industries that belong to the original set of domestic industries in 2006 as a percentage of the total amount of domestic industries in each consecutive year. The dotted line shows the complement, i.e., the share of domestic industries in each preceding year that still existed in 2019.}
    \label{Fig:StrucChange}
\end{figure}

\section{Results}\label{Sec:Results2} 

\subsection{Descriptive analysis}

As a preliminary step, we first investigate the magnitude of domestic regional structural change following the approach of \citet{neffke2011regions}. Figure~\ref{Fig:StrucChange} shows the dynamics of domestic industries over the entire period in our study. Regarding all industry-region combinations within our dataset, only 87\% present within 2006 still exists in 2019. Taking the reverse perspective, 85\% of domestic industries in 2019, already existed in 2006. For comparison, \citet{neffke2011regions} found that 78\% of industries present within 1998 in Sweden were still present in 2002 and 68\% of industries present in 2002 were still present in 1998. Slightly lower levels of churn are not unexpected in our case as we consider just a subset of Irish firms (i.e., those supported by government agencies). 

\begin{figure}[t!]
    \centering
    \includegraphics[width = 0.7\textwidth]{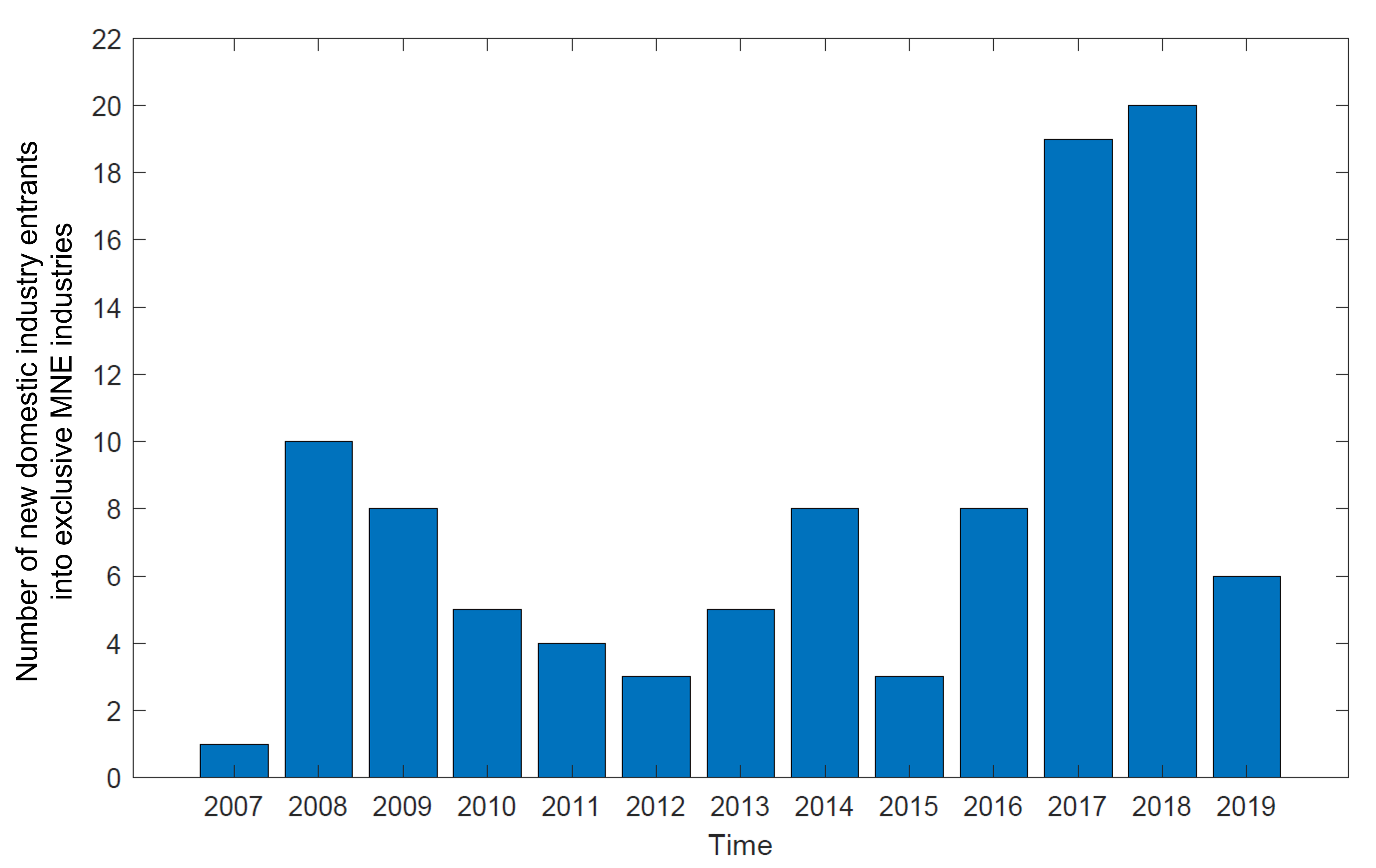}
    \caption{The number of new domestic industry entries into industries in which only MNEs are active in all regions in Ireland within the 2006-2019 period.}
    \label{FigDom2MNE}
\end{figure}

In Figure~\ref{FigDom2MNE}, we illustrate the number of new domestic industries entering into exclusive MNE industries within each year\footnote{In Appendix \ref{Appendix_2} we show that there is an increase in entries across all regions which is inline with the policy aim of increasing growth in regions particularly outside of Dublin \citep{MarketDiscoveryFund}}. We observe a sharp increase in entries within the 2017 and 2018 period. In 2017, due to high levels of uncertainty and fear of loss of UK markets by Irish exporting firms, the Irish government made a large amount of capital available to Irish firms, particularly SMEs, supported by government agencies \citep{Budget}. This investment and a range of corresponding policies aimed to both provide adequate support to Irish exporting firms \citep{IrishTimes2}, and enhance the diversification of export markets and promote domestic entry into existing markets \citep{MarketDiscoveryFund, IrishTimes}. 

In Table~\ref{Tab:DescMod} we show the descriptors of our dependent and explanatory variables. We also show the pairwise correlation between the various cohesion measures and the domestic industry entry and exit variables in Table~\ref{Tab:PairwiseCorr}. We see a positive but small correlation between the entry of domestic industries and the various cohesion measures. In accordance with the literature, this suggests that the more cohesive an industry the higher the likelihood of its entrance. On the other hand we see a negative relationship between the exit of domestic industries and the various cohesion measures. Once again, this agrees with the dominant view in the literature and suggests that the less cohesive an industry the higher its chance of exit. We now further investigate these relationships controlling for various effects using econometric models.

\begin{table}[b!]
\centering
\caption{Panel descriptors of dependent and explanatory variables}
\resizebox{0.5\textwidth}{!}{%
\begin{tabular}{l|lllll}
\hline
Variable & N & Mean & SD & Min & Max \\
\hline
$Entry_D$ & 10028 & 0.0363 & 0.1870 & 0 & 1 \\
$Exit_D$ & 11064 & 0.0287 & 0.1668 & 0 & 1 \\
$WC_{excl D}$ & 51632 & 3.4345 & 3.9097 & 0 & 29.5297 \\
$WC_{excl M}$ & 51632 & 0.4878 & 0.9065 & 0 & 10.1561 \\
$WC_{overlap}$ & 51632 & 1.3554 & 2.3019 & 0 & 21.2523 \\
$SC_{excl D}$ & 51632 & 0.0022 & 0.0021 & 0 & 0.0143 \\
$SC_{excl_M}$ & 51632 & 0.0022 & 0.0027 & 0 & 0.0239 \\
$SC_{overlap}$ & 51632 & 0.0022 & 0.0029 & 0 & 0.0245 \\
\hline
\end{tabular}}
\label{Tab:DescMod}
\end{table}

\begin{table}[t!]
\caption{Pairwise correlation of independent variables and explanatory variables.}
\resizebox{\textwidth}{!}{%
\begin{tabular}{l|lllllllll}
\hline
 & Entry\_D & Exit\_D & MNE Presence & $WC_{excl D}$ & $WC_{excl M}$ & $WC_{overlap}$ & $SC_{excl D}$ & $SC_{excl M}$ & $SC_{overlap}$ \\
 \hline
MNE Presence & \cellcolor[HTML]{6DA5CC}0.0906 & \cellcolor[HTML]{DC3D2D}-0.0728 & 1 &  &  &  &  &  &  \\
$WC_{excl D}$ & \cellcolor[HTML]{6DA5CC}0.0505 & \cellcolor[HTML]{DC3D2D}-0.0827 & 0.0043 & 1 &  &  &  &  &  \\
$WC_{excl M}$ & \cellcolor[HTML]{6DA5CC}0.0668 & \cellcolor[HTML]{DC3D2D}-0.0575 & 0.3043 & 0.436 & 1 &  &  &   &  \\
$WC_{overlap}$ & \cellcolor[HTML]{6DA5CC}0.0652 & \cellcolor[HTML]{DC3D2D}-0.0897 & 0.1609 & 0.448 & 0.421 & 1 &  &  &  \\
$SC_{excl D}$ & \cellcolor[HTML]{6DA5CC}0.0579 & \cellcolor[HTML]{DC3D2D}-0.0935 & 0.0278 & 0.6959 & 0.5144 & 0.5886 & 1 &  &  \\
$SC_{excl M}$ & \cellcolor[HTML]{6DA5CC}0.0563 & \cellcolor[HTML]{DC3D2D}-0.0344 & 0.1436 & 0.3783 & 0.5903 & 0.5625 & 0.4014 & 1 &  \\
$SC_{overlap}$ & \cellcolor[HTML]{6DA5CC}0.0698 & \cellcolor[HTML]{DC3D2D}-0.0915 & 0.1601 & 0.4530 & 0.5021 & 0.6953 & 0.5098 & 0.4961 & 1 \\
\hline
\end{tabular}
}
\label{Tab:PairwiseCorr} 
\\
\begin{small}
Note: Correlation values displayed in blocks of blue only include observations within Model 1 (as in Table~\ref{Tab:IndEntry_RI} and Table~\ref{Tab:IndEntry_SC}). Correlation values displayed in blocks of red only include observations within Model 2 (as in Table~\ref{Tab:IndExit_RI}, Table~\ref{Tab:IndExit_SC}). Correlation values displayed in blocks of white include all values with different industry-region-time combinations.
\end{small}
\end{table}

\subsection{Domestic industry entrance} \label{Sec_Results3}

The results of our econometric model in Equation~\ref{probit1} are reported in Table~\ref{Tab:IndEntry_RI} and Table~\ref{Tab:IndEntry_SC}. Each table is also sub-divided into three sections horizontally representing the three time periods we investigate independently. 

First, we focus on the relationship between local domestic activity and new domestic activity. Recall that the consensus within the regional branching literature is that the presence of related industries enhances industry entry. This is as regions grow by building on existing expertise and fostering new economic activities in related industries \citep{frenken2007related, neffke2011regions, boschma2014regional}. 

Consistent with the literature, for all three periods we find a positive and significant relationship for \textit{strategic} closeness to exclusive domestic industries which are dominated by low tech manufacturing and agriculture-related sectors. Hence, cohesion to deeply embedded domestic industries is associated with a higher probability of a domestic industry entry. This result highlights the important role of dense linkages \textit{between} sectors, and particularly domestic-only sectors, in enhancing the industrial diversification potential of a region.  

For the 2006-2009 and 2015-19 non-recession periods we observe a positive and significant relationship between new domestic activity and cohesion to overlapping industries (those in which both MNE and domestic firms are active), including both weighted and strategic closeness in the first period, and strategic closeness in the second. These industries are the more complex industries within a region's industrial basket, mostly consisting of medium-high tech manufacturing, information and telecommunication as well as professional service activities. Well-known examples include Ireland’s world-renowned baby food sector, which includes both domestic and foreign-owned global industry leaders. In particular, powdered milk has grown rapidly due to Ireland’s large dairy sector. It appears that domestic firms tend to enter new industries proximate to existing dynamic sectors, already home to a mix of domestic and foreign-owned enterprises. 

We observe the opposite effect when considering entries into industries proximate to MNE exclusive sectors. Specifically, for the 2006-2009 period, we find that a domestic firm is unlikely to enter an industry that is closely related to MNE-only industries. These industries are highly complex and less related to the region's domestic skill-base. For domestic industries to enter MNE-only industries (or those proximate to them) they need to make large cognitive 'leaps' to bridge the capability gap. 

Although we cannot disentangle the exact reasons why domestic firms are failing to enter industries linked to complex MNE-dominated industries, our findings very much relate to those on firm absorptive capacity \cite{kokko1996local,  blomstrom1998multinational}, also thought to be factor in Irish domestic firm productivity \citep{barrios2004efficiency, di2018productivity} (absorptive capacity is proxied in both cases by the presence of R\&D activity). Other possible reasons include: a potential lack of appropriate training and skill development of Irish workers to be able to work and learn from more complex industries dominated by MNEs, a lack of incentives for MNEs to engage in R\&D collaboration with domestic firms \citep{mcguirk2015measuring}, as well as potentially less strategic investment decisions by government agencies to encourage domestic firms to enter into MNE markets \citep{conefrey2018modelling, cortinovis_multinational_2020}.

During the crisis period, there is no significant relationship between industry entrance and cohesion to overlapping or exclusive MNE industries. It appears that entries into industries characterised by relatedness to complex sectors ceased during this difficult period. 

What is particularly striking about the recovery period is the strong significant and positive relationship between domestic industry entrance into MNE-only industries. Our analysis is picking up the Irish response to Brexit. In particular, we find that the dramatic level of government support and capital provided to Irish firms during 2017-18 appears to significantly reduce the previous barrier to entry of domestic industries into exclusive MNE industries. 

We show results combining $WC$ and $SC$ in a single model in Table~\ref{Tab:IndEntry_SC2} in Appendix~\ref{Appendix_3}. We generally find that our results hold. Our cohesion measures remain significant when controlling for the other cohesion measure, demonstrating empirically that these measures pick up different dimensions of cohesion. 

Overall, we observe that strategic closeness to exclusive domestic industries appears to be more important than simple weighted closeness. In terms of MNEs, both before and after the financial crisis cohesion to overlapping industries is associated with a higher probability of entry. In contrast, cohesion to MNE-only industries is associated with a lower probability of entry pre-crisis. Post crisis, however, Brexit funds appear to induce entry into MNE exclusive industries.  

\subsection{Domestic industry exit} \label{Sec_Results4}

The results of our second econometric model in Equation~\ref{probit2} are shown in Table~\ref{Tab:IndExit_RI}, and Table~\ref{Tab:IndExit_SC} for the cohesion measures $WC$ and $SC$, respectively. 

Recall that within both the regional branching and resilience literature it is generally accepted that industries which are less related to a region's industrial basket are more likely to exit \citep{neffke2011regions}. Hence, relatedness amongst industries provides a protective shield against industry exit in response to economic shocks \citep{essletzbichler_relatedness_2015, balland_technological_2015}.

In agreement with the literature, we find that industry cohesion to exclusively domestic industries has a protective effect. Specifically, cohesion to `strategic' domestic industries enhanced industry survival in the latter two periods. The protective role of related strategic industries during the recession period, a time when industry exits peaked, is particularly important. The presence of higher-order connections between domestic industries in the region implies a cluster or concentration of related activities, which gives rise to a wider range of opportunities for beneficial linkages and spillovers. 

Similar to entries, overlapping industries also play a key role in domestic industry survival. In particular, we find a negative and significant relationship for weighted closeness with domestic industry exit for the first two periods (pre-recession and recession), and strategic closeness during the first period. Hence, the more cohesive an industry is to these dynamic industries, particularly during the recession, the more likely it is to survive. 

In the 2006-2009 period, we see a significant and negative effect of the presence of MNEs within an industry and its exit. Hence, having MNEs active within an industry (and therefore being an overlapping industry) enhances resilience - but only in the pre-recession period. This corresponds with the findings of \citet{gorg2003multinational} who find a positive impact of MNEs on domestic manufacturing plants in high tech sectors between 1973-96. In the recession period we observed a significant and positive effect of both weighted and strategic closeness, suggesting that industries that were related to MNE-only industries were more likely to exit. It is likely that these industries experienced weaker ties to both domestic and MNE firms in their region, and were hence more likely to exit under duress. 

We also show the results for the $SC$ variable when controlling for the corresponding $WC$ measure in Table~\ref{Tab_IndExits_SC2} in Appendix~\ref{Appendix_3}. As before, our cohesion measures remain significant when controlling for the other cohesion measure.

Overall, it appears that cohesion to overlapping industries as well as strategic domestic industries have a protective effect in a crisis. Cohesion to exclusive MNE industries, however, was associated with an increased probability of exit during this period. 

\begin{landscape}

\begin{table}
\caption{Panel probit regression results for domestic industry entrance between 2006-2019 and their weighed closeness cohesive measure as independent variable}
\resizebox{1.5\textwidth}{!}{%
\begin{tabular}{l|lllll|lllll|lllll}
\hline
Baseline period & \multicolumn{5}{c|}{2006-2009} & \multicolumn{5}{c|}{2010-2014} & \multicolumn{5}{c|}{2015-2019} \\ \hline
 & \multicolumn{1}{c}{(1)} & \multicolumn{1}{c}{(2)} & \multicolumn{1}{c}{(3)} & \multicolumn{1}{c}{(4)} & \multicolumn{1}{c|}{(5)} & \multicolumn{1}{c}{(1)} & \multicolumn{1}{c}{(2)} & \multicolumn{1}{c}{(3)} & \multicolumn{1}{c}{(4)} & \multicolumn{1}{c|}{(5)} & \multicolumn{1}{c}{(1)} & \multicolumn{1}{c}{(2)} & \multicolumn{1}{c}{(3)} & \multicolumn{1}{c}{(4)} & \multicolumn{1}{c}{(5)} \\ \hline
MNE industry & \begin{tabular}[c]{@{}l@{}}-0.439\\ (0.315)\end{tabular} & \begin{tabular}[c]{@{}l@{}}-0.433\\ (0.315)\end{tabular} & \begin{tabular}[c]{@{}l@{}}-0.477\\ (0.319)\end{tabular} & \begin{tabular}[c]{@{}l@{}}-0.568 \\ (0.332)\end{tabular} & \begin{tabular}[c]{@{}l@{}}-0.578\\ (0.333)\end{tabular} & \begin{tabular}[c]{@{}l@{}}0.021\\ (0.306)\end{tabular} & \begin{tabular}[c]{@{}l@{}}-0.024 \\ (0.306)\end{tabular} & \begin{tabular}[c]{@{}l@{}}0.048 \\ (0.308)\end{tabular} & \begin{tabular}[c]{@{}l@{}}0.020 \\ (0.306)\end{tabular} & \begin{tabular}[c]{@{}l@{}}0.011 \\ (0.309)\end{tabular} & \begin{tabular}[c]{@{}l@{}}1.673*** \\ (0.451)\end{tabular} & \begin{tabular}[c]{@{}l@{}}1.642*** \\ (0.453)\end{tabular} & \begin{tabular}[c]{@{}l@{}}1.690*** \\ (0.457)\end{tabular} & \begin{tabular}[c]{@{}l@{}}1.715***\\ (0.465)\end{tabular} & \begin{tabular}[c]{@{}l@{}}1.673*** \\ (0.470)\end{tabular} \\
$WC_{excl D}$ &  & \begin{tabular}[c]{@{}l@{}}0.047\\ (0.054)\end{tabular} &  &  & \begin{tabular}[c]{@{}l@{}}0.047\\ (0.056)\end{tabular} &  & \begin{tabular}[c]{@{}l@{}}0.132*\\ (0.076)\end{tabular} &  &  & \begin{tabular}[c]{@{}l@{}}0.134*\\ (0.078)\end{tabular} &  & \begin{tabular}[c]{@{}l@{}}0.065 \\ (0.100)\end{tabular} &  &  & \begin{tabular}[c]{@{}l@{}}0.106 \\ (0.107)\end{tabular} \\
$WC_{excl M}$ &  &  & \begin{tabular}[c]{@{}l@{}}-0.259**\\ (0.138)\end{tabular} &  & \begin{tabular}[c]{@{}l@{}}-0.185* \\ (-0.146)\end{tabular} &  &  & \begin{tabular}[c]{@{}l@{}}0.209 \\ (0.145)\end{tabular} &  & \begin{tabular}[c]{@{}l@{}}0.246\\ (0.165)\end{tabular} &  &  & \begin{tabular}[c]{@{}l@{}}0.169 \\ (0.218)\end{tabular} &  & \begin{tabular}[c]{@{}l@{}}0.11 \\ (0.229)\end{tabular} \\
$WC_{overlap}$ &  &  &  & \begin{tabular}[c]{@{}l@{}}0.231*** \\ (0.072)\end{tabular} & \begin{tabular}[c]{@{}l@{}}0.222*** \\ (0.074)\end{tabular} &  &  &  & \begin{tabular}[c]{@{}l@{}}0.020\\ (0.075)\end{tabular} & \begin{tabular}[c]{@{}l@{}}0.066\\ (0.089)\end{tabular} &  &  &  & \begin{tabular}[c]{@{}l@{}}0.168 \\ (0.107)\end{tabular} & \begin{tabular}[c]{@{}l@{}}0.186 \\ (0.115)\end{tabular} \\
Region FE & Y & Y & Y & Y & Y & Y & Y & Y & Y & Y & Y & Y & Y & Y & Y \\
Industry FE & Y & Y & Y & Y & Y & Y & Y & Y & Y & Y & Y & Y & Y & Y & Y \\
Constant & \begin{tabular}[c]{@{}l@{}}-8.841\\ (8.14e+06)\end{tabular} & \begin{tabular}[c]{@{}l@{}}-8.405\\ (4.97e+06)\end{tabular} & \begin{tabular}[c]{@{}l@{}}-8.302\\ (8.14e+06)\end{tabular} & \begin{tabular}[c]{@{}l@{}}-8.323 \\ (2.85e+06)\end{tabular} & \begin{tabular}[c]{@{}l@{}}-8.590\\ (4.21e+06)\end{tabular} & \begin{tabular}[c]{@{}l@{}}-42.689\\ (5.76e+06)\end{tabular} & \begin{tabular}[c]{@{}l@{}}-9.867\\ (5.76e+06)\end{tabular} & \begin{tabular}[c]{@{}l@{}}-19.189\\ (5.76e+06)\end{tabular} & \begin{tabular}[c]{@{}l@{}}-25.688 \\ (5.75e+06)\end{tabular} & \begin{tabular}[c]{@{}l@{}}-19.177 \\ (5.75e+06)\end{tabular} & \begin{tabular}[c]{@{}l@{}}-16.145\\ (5.76e+06)\end{tabular} & \begin{tabular}[c]{@{}l@{}}-17.634\\ (5.75e+06)\end{tabular} & \begin{tabular}[c]{@{}l@{}}-15.652 \\ (5.75e+06)\end{tabular} & \begin{tabular}[c]{@{}l@{}}-18.6160 \\ (5.75e+06)\end{tabular} & \begin{tabular}[c]{@{}l@{}}-14.952 \\ (5.75e+06)\end{tabular} \\ \hline
N & 2522 & 2522 & 2522 & 2522 & 2522 & 2494 & 2494 & 2494 & 2494 & 2494 & 2507 & 2507 & 2507 & 2507 & 2507 \\
AUC & 0.9480 & 0.9481 & 0.9480 & 0.9508 & 0.9509 & 0.9700 & 0.9703 & 0.9706 & 0.9700 & 0.9710 & 0.9814 & 0.9811 & 0.9819 & 0.9819 & 0.9821 \\ \hline
\end{tabular}}
\small {Notes: Robust standard error in parenthesis; * p<0.1, ** p< 0.05, *p<0.01}
\label{Tab:IndEntry_RI}
\end{table}

\begin{table}
\caption{Panel probit regression results for domestic industry entrance between 2006-2019 and their strategic closeness cohesive measure as independent variable}
\resizebox{1.5\textwidth}{!}{%
\begin{tabular}{l|lllll|lllll|lllll}
\hline
Baseline period & \multicolumn{5}{c|}{2006-2009} & \multicolumn{5}{c|}{2010-2014} & \multicolumn{5}{c|}{2015-2019} \\ \hline
 & \multicolumn{1}{c}{(1)} & \multicolumn{1}{c}{(2)} & \multicolumn{1}{c}{(3)} & \multicolumn{1}{c}{(4)} & \multicolumn{1}{c|}{(5)} & \multicolumn{1}{c}{(1)} & \multicolumn{1}{c}{(2)} & \multicolumn{1}{c}{(3)} & \multicolumn{1}{c}{(4)} & \multicolumn{1}{c|}{(5)} & \multicolumn{1}{c}{(1)} & \multicolumn{1}{c}{(2)} & \multicolumn{1}{c}{(3)} & \multicolumn{1}{c}{(4)} & \multicolumn{1}{c}{(5)} \\ \hline
MNE industry & \begin{tabular}[c]{@{}l@{}}-0.439 \\ (0.315)\end{tabular} & \begin{tabular}[c]{@{}l@{}}-0.474 \\ (0.317)\end{tabular} & \begin{tabular}[c]{@{}l@{}}-0.253\\ (0.351)\end{tabular} & \begin{tabular}[c]{@{}l@{}}-0.547\\ (0.323)\end{tabular} & \begin{tabular}[c]{@{}l@{}}-0.399\\ (0.360)\end{tabular} & \begin{tabular}[c]{@{}l@{}}0.021 \\ (0.306)\end{tabular} & \begin{tabular}[c]{@{}l@{}}0.008\\ (0.305)\end{tabular} & \begin{tabular}[c]{@{}l@{}}0.098\\ (0.331)\end{tabular} & \begin{tabular}[c]{@{}l@{}}0.027 \\ (0.306)\end{tabular} & \begin{tabular}[c]{@{}l@{}}0.0823\\ (0.330)\end{tabular} &  \begin{tabular}[c]{@{}l@{}}1.673*** \\ (0.451)\end{tabular} & \begin{tabular}[c]{@{}l@{}}1.666***\\ (0.451)\end{tabular} & \begin{tabular}[c]{@{}l@{}}1.716***\\ (0.496)\end{tabular} & \begin{tabular}[c]{@{}l@{}}1.6722*** \\ (0.453)\end{tabular} & \begin{tabular}[c]{@{}l@{}}1.677*** \\ (0.499)\end{tabular} \\
$SC_{excl D}$ &  & 302.29** &  &  & \begin{tabular}[c]{@{}l@{}}262.48** \\ (115.86)\end{tabular} &  & \begin{tabular}[c]{@{}l@{}}283.89*\\ (119.36)\end{tabular} &  &  & \begin{tabular}[c]{@{}l@{}}319.41*\\ (203.4)\end{tabular} &  & \begin{tabular}[c]{@{}l@{}}148** \\ (195.97)\end{tabular} &  &  & \begin{tabular}[c]{@{}l@{}}143.72** \\ (197.84)\end{tabular} \\
$SC_{excl M}$ &  &  & \begin{tabular}[c]{@{}l@{}}-65.379\\ (54.879)\end{tabular} &  & \begin{tabular}[c]{@{}l@{}}-67.676\\ (55.878)\end{tabular} &  &  & \begin{tabular}[c]{@{}l@{}}-37.703\\ (60.961)\end{tabular} &  & \begin{tabular}[c]{@{}l@{}}-43.182\\ (62.164)\end{tabular} &  &  & \begin{tabular}[c]{@{}l@{}}-18.482 \\ (84.414)\end{tabular} &  & \begin{tabular}[c]{@{}l@{}}-5.736 \\ (57.497)\end{tabular} \\
$SC_{overlap}$ &  &  &  & \begin{tabular}[c]{@{}l@{}}116.46**\\ (63.948)\end{tabular} & \begin{tabular}[c]{@{}l@{}}104.66* \\ (66.002)\end{tabular} &  &  &  & \begin{tabular}[c]{@{}l@{}}33.677\\ (69.689)\end{tabular} & \begin{tabular}[c]{@{}l@{}}42.581\\ (70.587)\end{tabular} &  &  &  & \begin{tabular}[c]{@{}l@{}}73.628* \\ (110.78)\end{tabular} & \begin{tabular}[c]{@{}l@{}}70.852*\\ (111.67)\end{tabular} \\
Region FE & Y & Y & Y & Y & Y & Y & Y & Y & Y & Y & Y & Y & Y & Y & Y \\
Industry FE & Y & Y & Y & Y & Y & Y & Y & Y & Y & Y & Y & Y & Y & Y & Y \\
Constant & \begin{tabular}[c]{@{}l@{}}-8.841 \\ (8.14e+06)\end{tabular} & \begin{tabular}[c]{@{}l@{}}-9.741 \\ (3.53e+06)\end{tabular} & \begin{tabular}[c]{@{}l@{}}-7.836\\ (6.57e+06)\end{tabular} & \begin{tabular}[c]{@{}l@{}}-9.431\\ (8.13e+06)\end{tabular} & \begin{tabular}[c]{@{}l@{}}-9.932\\ (3.93e+06)\end{tabular} & \begin{tabular}[c]{@{}l@{}}-42.689\\ (5.76e+06)\end{tabular} & \begin{tabular}[c]{@{}l@{}}-22.436 \\ (5.75e+06)\end{tabular} & \begin{tabular}[c]{@{}l@{}}-27.531\\ (5.75e+06)\end{tabular} & \begin{tabular}[c]{@{}l@{}}-23.761\\ (5.75e+06)\end{tabular} & \begin{tabular}[c]{@{}l@{}}-12.78\\ (5.75e+06)\end{tabular} & \begin{tabular}[c]{@{}l@{}}-16.145 \\ (5.75e+06)\end{tabular} & \begin{tabular}[c]{@{}l@{}}-24.037 \\ (5.75e+06)\end{tabular} & \begin{tabular}[c]{@{}l@{}}-16.396 \\ (5.75e+06)\end{tabular} & \begin{tabular}[c]{@{}l@{}}-20.035\\ (5.75e+06)\end{tabular} & \begin{tabular}[c]{@{}l@{}}-18.247 \\ (5.75e+06)\end{tabular} \\ \hline
N & 2522 & 2522 & 2522 & 2522 & 2522 & 2494 & 2494 & 2494 & 2494 & 2494 & 2507 & 2507 & 2507 & 2507 & 2507 \\
AUC & 0.9480 & 0.9487 & 0.9474 & 0.9487 & 0.9490 & 0.9700 & 0.9706 & 0.9010 & 0.9700 & 0.9707 & 0.9814 & 0.9813 & 0.9813 & 0.9816 & 0.9817 \\ \hline
\end{tabular}
}
\small {Notes: Robust standard error in parenthesis; * p<0.1, ** p< 0.05, *p<0.01}
\label{Tab:IndEntry_SC}
\end{table}

\begin{table}
\caption{Panel probit regression results for domestic industry exits between 2006-2019 and their weighted closeness cohesive measure as independent variable}
\resizebox{1.5\textwidth}{!}{%
\begin{tabular}{l|lllll|lllll|lllll}
\hline
Baseline period & \multicolumn{5}{c|}{2006-2009} & \multicolumn{5}{c|}{2010-2014} & \multicolumn{5}{c|}{2015-2019} \\ \hline
 & \multicolumn{1}{c}{(1)} & \multicolumn{1}{c}{(2)} & \multicolumn{1}{c}{(3)} & \multicolumn{1}{c}{(4)} & \multicolumn{1}{c|}{(5)} & \multicolumn{1}{c}{(1)} & \multicolumn{1}{c}{(2)} & \multicolumn{1}{c}{(3)} & \multicolumn{1}{c}{(4)} & \multicolumn{1}{c|}{(5)} & \multicolumn{1}{c}{(1)} & \multicolumn{1}{c}{(2)} & \multicolumn{1}{c}{(3)} & \multicolumn{1}{c}{(4)} & \multicolumn{1}{c}{(5)} \\ \hline
MNE industry & \begin{tabular}[c]{@{}l@{}}-0.607**\\ (0.271)\end{tabular} & \begin{tabular}[c]{@{}l@{}}-0.570**\\ (0.281)\end{tabular} & \begin{tabular}[c]{@{}l@{}}-0.593**\\ (0.272)\end{tabular} & \begin{tabular}[c]{@{}l@{}}-0.586**\\ (0.274)\end{tabular} & \begin{tabular}[c]{@{}l@{}}-0.569** \\ (0.286)\end{tabular} & \begin{tabular}[c]{@{}l@{}}-0.140 \\ (0.255)\end{tabular} & \begin{tabular}[c]{@{}l@{}}-0.140 \\ (0.255)\end{tabular} & \begin{tabular}[c]{@{}l@{}}-0.118 \\ (0.257)\end{tabular} & \begin{tabular}[c]{@{}l@{}}-0.036\\ (0.264)\end{tabular} & \begin{tabular}[c]{@{}l@{}}-0.048 \\ (0.265)\end{tabular} & \begin{tabular}[c]{@{}l@{}}0.025\\ (0.334)\end{tabular} & \begin{tabular}[c]{@{}l@{}}0.028 \\ (0.337)\end{tabular} & \begin{tabular}[c]{@{}l@{}}0.028\\ (0.335)\end{tabular} & \begin{tabular}[c]{@{}l@{}}0.040 \\ (0.334)\end{tabular} & \begin{tabular}[c]{@{}l@{}}0.114\\ (0.339)\end{tabular} \\
$WC_{excl D}$ &  & \begin{tabular}[c]{@{}l@{}}-0.304***\\ (0.097)\end{tabular} &  &  & \begin{tabular}[c]{@{}l@{}}-0.337***\\ (0.102)\end{tabular} &  & \begin{tabular}[c]{@{}l@{}}-0.068\\ (0.069)\end{tabular} &  &  & \begin{tabular}[c]{@{}l@{}}-0.103 \\ (0.077)\end{tabular} &  & \begin{tabular}[c]{@{}l@{}}-0.146\\ (0.086)\end{tabular} &  &  & \begin{tabular}[c]{@{}l@{}}-0.128 \\ (0.090)\end{tabular} \\
$WC_{excl M}$ &  &  & \begin{tabular}[c]{@{}l@{}}0.094\\ (0.150)\end{tabular} &  & \begin{tabular}[c]{@{}l@{}}-0.049\\ (0.163)\end{tabular} &  &  & \begin{tabular}[c]{@{}l@{}}0.340*** \\ (0.166)\end{tabular} &  & \begin{tabular}[c]{@{}l@{}}0.235**\\ (0.191)\end{tabular} &  &  & \begin{tabular}[c]{@{}l@{}}-0.726 \\ (0.169)\end{tabular} &  & \begin{tabular}[c]{@{}l@{}}-0.269\\ (0.173)\end{tabular} \\
$WC_{overlap}$ &  &  &  & \begin{tabular}[c]{@{}l@{}}-0.148** \\ (0.084)\end{tabular} & \begin{tabular}[c]{@{}l@{}}-0.204**\\ (0.102)\end{tabular} &  &  &  & \begin{tabular}[c]{@{}l@{}}-0.372*** \\ (0.098)\end{tabular} & \begin{tabular}[c]{@{}l@{}}-0.372***\\ (0.111)\end{tabular} &  &  &  & \begin{tabular}[c]{@{}l@{}}-0.093 \\ (0.335)\end{tabular} & \begin{tabular}[c]{@{}l@{}}-0.105 \\ (0.078)\end{tabular} \\
Region FE & Y & Y & Y & Y & Y & Y & Y & Y & Y & Y & Y & Y & Y & Y & Y \\
Industry FE & Y & Y & Y & Y & Y & Y & Y & Y & Y & Y & Y & Y & Y & Y & Y \\
Constant & \begin{tabular}[c]{@{}l@{}}-1.193*\\ (0.622)\end{tabular} & \begin{tabular}[c]{@{}l@{}}1.165\\ (0.979)\end{tabular} & \begin{tabular}[c]{@{}l@{}}-1.207* \\ (0.623)\end{tabular} & \begin{tabular}[c]{@{}l@{}}-1.030\\ (0.641)\end{tabular} & \begin{tabular}[c]{@{}l@{}}1.623\\ (1.036)\end{tabular} & \begin{tabular}[c]{@{}l@{}}-21.305 \\ (3.32e+06)\end{tabular} & \begin{tabular}[c]{@{}l@{}}-14.167 \\ (3.32e+06)\end{tabular} & \begin{tabular}[c]{@{}l@{}}-10.715\\ (3.32e+06)\end{tabular} & \begin{tabular}[c]{@{}l@{}}-16.713 \\ (3.32e+06)\end{tabular} & \begin{tabular}[c]{@{}l@{}}-16.943 \\ (3.32e+06)\end{tabular} & \begin{tabular}[c]{@{}l@{}}-15.163 \\ (3.32e+06)\end{tabular} & \begin{tabular}[c]{@{}l@{}}-12.568 \\ (3.32e+06)\end{tabular} & \begin{tabular}[c]{@{}l@{}}-14.402 \\ (3.32e+06)\end{tabular} & \begin{tabular}[c]{@{}l@{}}-14.734 \\ (3.32e+06)\end{tabular} & \begin{tabular}[c]{@{}l@{}}-15.568 \\ (3.32e+06)\end{tabular} \\ \hline
N & 1166 & 1166 & 1166 & 1166 & 1166 & 1194 & 1194 & 1194 & 1194 & 1194 & 1181 & 1181 & 1181 & 1181 & 1181 \\
AUC & 0.9466 & 0.9512 & 0.9460 & 0.9455 & 0.9514 & 0.9479 & 0.9478 & 0.9501 & 0.9549 & 0.9561 & 0.9650 & 0.9657 & 0.9656 & 0.9645 & 0.9662 \\ \hline
\end{tabular}}
\small {Notes: Robust standard error in parenthesis; * p<0.1, ** p< 0.05, *p<0.01}
\label{Tab:IndExit_RI}
\end{table}

\begin{table}
\caption{Panel probit regression results for domestic industry exit between 2006-2019 and their strategic closeness cohesive measure as independent variable}
\resizebox{1.5\textwidth}{!}{%
\begin{tabular}{l|lllll|lllll|lllll}
\hline
Baseline period & \multicolumn{5}{c|}{2006-2009} & \multicolumn{5}{c|}{2010-2014} & \multicolumn{5}{c|}{2015-2019} \\ \hline
 & \multicolumn{1}{c}{(1)} & \multicolumn{1}{c}{(2)} & \multicolumn{1}{c}{(3)} & \multicolumn{1}{c}{(4)} & \multicolumn{1}{c|}{(5)} & \multicolumn{1}{c}{(1)} & \multicolumn{1}{c}{(2)} & \multicolumn{1}{c}{(3)} & \multicolumn{1}{c}{(4)} & \multicolumn{1}{c|}{(5)} & \multicolumn{1}{c}{(1)} & \multicolumn{1}{c}{(2)} & \multicolumn{1}{c}{(3)} & \multicolumn{1}{c}{(4)} & \multicolumn{1}{c}{(5)} \\ \hline
MNE industry & \begin{tabular}[c]{@{}l@{}}-0.606** \\ (0.271)\end{tabular} & \begin{tabular}[c]{@{}l@{}}-0.711** \\ (0.281)\end{tabular} & \begin{tabular}[c]{@{}l@{}}-0.607** \\ (0.272)\end{tabular} & \begin{tabular}[c]{@{}l@{}}-0.607**\\ (0.272)\end{tabular} & \begin{tabular}[c]{@{}l@{}}-0.546*\\ (0.306)\end{tabular} & \begin{tabular}[c]{@{}l@{}}-0.140\\ (0.255)\end{tabular} & \begin{tabular}[c]{@{}l@{}}-0.223 \\ (0.262)\end{tabular} & \begin{tabular}[c]{@{}l@{}}-0.234 \\ (0.263)\end{tabular} & \begin{tabular}[c]{@{}l@{}}-0.234 \\ (0.280)\end{tabular} & \begin{tabular}[c]{@{}l@{}}-0.193\\ (0.301)\end{tabular} & \begin{tabular}[c]{@{}l@{}}0.0248 \\ (0.334)\end{tabular} & \begin{tabular}[c]{@{}l@{}}0.010\\ (0.338)\end{tabular} & \begin{tabular}[c]{@{}l@{}}0.024 \\ (0.334)\end{tabular} & \begin{tabular}[c]{@{}l@{}}0.1195 \\ (0.343)\end{tabular} & \begin{tabular}[c]{@{}l@{}}0.078\\ (0.348)\end{tabular} \\
$SC_{excl D}$ &  & \begin{tabular}[c]{@{}l@{}}-377 \\ (240.64)\end{tabular} &  &  & \begin{tabular}[c]{@{}l@{}}-199.77\\ (255.3)\end{tabular} &  & \begin{tabular}[c]{@{}l@{}}-321.45*\\ (177.01)\end{tabular} &  &  & \begin{tabular}[c]{@{}l@{}}-277.44* \\ (183.96)\end{tabular} &  & \begin{tabular}[c]{@{}l@{}}-362.69*\\ (200.63)\end{tabular} &  &  & \begin{tabular}[c]{@{}l@{}}-323.78*\\ (203.92)\end{tabular} \\
$SC_{excl M}$ &  &  & \begin{tabular}[c]{@{}l@{}}104.81 \\ (73.94)\end{tabular} &  & \begin{tabular}[c]{@{}l@{}}103.54\\ (76.065)\end{tabular} &  &  & \begin{tabular}[c]{@{}l@{}}152.23** \\ (71.173)\end{tabular} &  & \begin{tabular}[c]{@{}l@{}}143.91** \\ (71.253)\end{tabular} &  &  & \begin{tabular}[c]{@{}l@{}}-10.645\\ (84.107)\end{tabular} &  & \begin{tabular}[c]{@{}l@{}}-13.787\\ (88.511)\end{tabular} \\
$SC_{overlap}$ &  &  &  & \begin{tabular}[c]{@{}l@{}}-204.81**\\ (98.647)\end{tabular} & \begin{tabular}[c]{@{}l@{}}-202.06**\\ (99.477)\end{tabular} &  &  &  & \begin{tabular}[c]{@{}l@{}}-124.84 \\ (87.733)\end{tabular} & \begin{tabular}[c]{@{}l@{}}-76.709 \\ (90.601)\end{tabular} &  &  &  & \begin{tabular}[c]{@{}l@{}}-139.28 \\ (995.135)\end{tabular} & \begin{tabular}[c]{@{}l@{}}-120.1\\ (0.218)\end{tabular} \\
Region FE & Y & Y & Y & Y & Y & Y & Y & Y & Y & Y & Y & Y & Y & Y & Y \\
Industry FE & Y & Y & Y & Y & Y & Y & Y & Y & Y & Y & Y & Y & Y & Y & Y \\
Constant & \begin{tabular}[c]{@{}l@{}}-1.193** \\ (0.622)\end{tabular} & \begin{tabular}[c]{@{}l@{}}-0.868 \\ (1.467)\end{tabular} & \begin{tabular}[c]{@{}l@{}}-1.428 \\ (1.847)\end{tabular} & \begin{tabular}[c]{@{}l@{}}-1.428\\ (1.847)\end{tabular} & \begin{tabular}[c]{@{}l@{}}-0.333\\ (1.533)\end{tabular} & \begin{tabular}[c]{@{}l@{}}-21.305\\ (3.32e+06)\end{tabular} & \begin{tabular}[c]{@{}l@{}}-12.079\\ (3.32e+06)\end{tabular} & \begin{tabular}[c]{@{}l@{}}-13.698\\ (3.32e+06)\end{tabular} & \begin{tabular}[c]{@{}l@{}}-14.851\\ (3.32e+06)\end{tabular} & \begin{tabular}[c]{@{}l@{}}-15.035\\ (3.32e+06)\end{tabular} & \begin{tabular}[c]{@{}l@{}}-15.163\\ (3.32e+06)\end{tabular} & \begin{tabular}[c]{@{}l@{}}-12.602\\ (3.32e+06)\end{tabular} & \begin{tabular}[c]{@{}l@{}}-14.076 \\ (3.32e+06)\end{tabular} & \begin{tabular}[c]{@{}l@{}}-16.625\\ (3.32e+06)\end{tabular} & \begin{tabular}[c]{@{}l@{}}-14.8\\ (3.32e+06)\end{tabular} \\ \hline
N & 1166 & 1166 & 1166 & 1166 & 1166 & 1194 & 1194 & 1194 & 1194 & 1194 & 1181 & 1181 & 1181 & 1181 & 1181 \\
AUC & 0.9466 & 0.9465 & 0.9466 & 0.9466 & 0.9491 & 0.9479 & 0.9484 & 0.9505 & 0.9486 & 0.9486 & 0.9650 & 0.9669 & 0.9651 & 0.9663 & 0.9671 \\ \hline
\end{tabular}
}
\small {Notes: Robust standard error in parenthesis; * p<0.1, ** p< 0.05, *p<0.01}
\label{Tab:IndExit_SC}
\end{table}

\end{landscape}


\section{Conclusions and policy implications}

It is well-established that regions grow by learning how to recombine complementary capabilities to move into more complex and sophisticated economic activities. The growth trajectory of a region is therefore conditioned on it's current capability base. With the rise of multinational enterprises globally, an issue of key concern is whether and how MNEs can act successfully as a channel to `import' new capabilities to a region and promote local domestic diversification.  

Focusing on a set of government-supported Irish firms active in manufacturing and exports, we find a strong role for so-called overlapping industries - those that have both domestic and MNE employment in a region. Specifically, domestic industries are both more likely to enter and less likely to leave a region if they are closely related to these industries. These are some of the most dynamic domestic sectors, home to global brands such as Glanbia, the Irish baby food manufacturer. While we cannot separate the role of MNEs from domestic firms within this subset of industries in terms of new entries, we can deduce that there exists a set of related sectors in which domestic firms already successfully thrive alongside MNEs, and it is this set that is driving new entries.   

In contrast to overlapping industries, we find a negative impact from 'exclusive MNE' industries. In particular, we find that cohesion of a domestic industry to MNE-only industries both reduces its chances of entry into a region as well as reduces its chance of survival in a crisis. These results suggest that domestic firms are unable to 'leap' into MNE-proximate industries, likely due to a technology or know-how gap that is too large to bridge. Furthermore, those that do are less likely to survive, suggesting weak ties. 

Finally, the type of cohesion matters. We differentiate between simple relatedness and strategic relatedness, or the presence of 'higher order' connections \textit{between} industries in a region. Our results show that dimension is particularly important for domestic industries: industry entries are associated with strategic rather than simple cohesion to both domestic and overlapping industries. Hence, entries tend to occur in industries proximate to regional 'clusters', or groups of existing interconnected industries. Similarly, exits are negatively correlated with strategic closeness to domestic and overlapping industries, further suggesting that embeddedness in a regional network of related industries is key to success. 

In previous work we identified two distinct export clusters in the Irish economy \citep{o2016tale}, visualised in the Product Space network of \citet{hidalgo2007product}. One of these clusters, located centrally in the network, contained industries with both Irish and MNE activities such as food and agriculture, while the other cluster was more peripheral, containing very complex sectors such as chemicals, pharmaceuticals and electronics. We hypothesised at the time that the ‘distance’ between these sectors would likely prevent domestic firms emerging in MNE sectors due to the huge capability gap. Our findings here very much accord with this hypothesis. Specifically, we find that the ‘rich get richer’ in the sense that overlapping industries in a region tend to attract further related industries. On the other hand, domestic entries into MNE dominated industries are not only rare but decrease in likelihood the closer the target industry is to MNE activities. This latter relationship is only ‘broken’ by a massive injection of funds in the wake of Brexit. 

There are a number of clear policy implications from our study. Firstly, our study suggests that industrial policy should distinguish types of MNEs, looking closely at sectoral concentrations and the potential for linkages to the domestic economy. Crucially, it suggests that policy should prioritise MNEs in overlapping industries, or those proximate to overlapping industries, as these have the greatest likelihood of inducing domestic transformation. This idea is somewhat at odds with a general strategy aimed at increasing the industrial or export complexity of a nation, or focusing on taxation income alone, irrespective of the domestic capability base. These findings are particularly salient within a context of finite investment resources and a potential global re-organisation of MNE activities resulting from international agreements on MNE taxation. 

The second policy relevant finding is that with large injections of funds, it is possible to induce domestic entry into MNE and MNE-related sectors. In the wider literature, this is known as leapfrogging, or unrelated diversification, and is known to be rare \cite{pinheiro2018shooting}. This is the first study that we know of that captures the impact of the Brexit ‘Market Discovery Fund’, which effectively doubled the size of the IDA (Investment and Development Agency) budget between 2016 and 2017, injecting massive capital into affected firms. Although beyond the scope and time-frame of this study, it would be interesting in future work to investigate the survival of these Brexit era entries to see whether their survival is comparable, better or worse than non Brexit induced entries. While more research into this policy is no doubt needed, our work would suggest that large targeted programmes backed by significant investment is needed to induce such leaps. 

Thirdly, in line with a burgeoning literature, methods from network science and data science can provide invaluable novel insights into the structure and dynamics of economic processes. In this case, we harness data on inter-sectoral labour mobility and network science to quantify the cohesion between industries. Establishing a mathematical model for cohesion, and differentiating types of cohesion, is key to predicting which industries are well-placed to enter a region, providing an evidence base for industrial policies such as grant funding, training and R\&D programmes, and infrastructure investment. Such modelling is rarely conclusive on its own, but forms part of a package of analysis which can be used to develop informed and strategic investments. 

Finally, we note that there are some limitations to this study. Most obviously, our data does not include services outside the exporting sector which are increasingly a large part of economic activity and employment both in Ireland and globally. Due to the knowledge intensive nature of these activities, we would expect that the relationships found here would both generalise and become stronger for service industries \citep{diodato2018industries} but this remains to be tested. Secondly, as discussed above, we are limited by the aggregate nature of our data, particularly in disentangling the influence of overlapping industries. Firm level data would enable a deeper analysis in future work. 

\clearpage
\begin{footnotesize}

\end{footnotesize}


\clearpage
\setcounter{section}{0}
\Large {\bf{Appendix}}
\normalsize {
\section{Data descriptive and matching}\label{Appendix_Data}

In this study we adopt two datasets. The first is derived from the Irish Department of Business, Enterprise, and Innovation Annual Business Survey of Economic Impact and covers a large subset of exporting firms within the Irish economy assisted by government agencies. The dataset contains the employment size of industries (broken down into 4-digit NACE $2$ industry codes) across regions (defined by NUTS III) and years. Furthermore, the dataset is broken down into employees who work for Irish firms and those who work for firms predominately owned through foreign investment. In table~\ref{Tab_DataDes} we show the number of MNE and domestic industries within each year and each Irish region that we use as observations within our analysis. We observe that there is a higher number of domestic industries compared to MNE industries in all regions and time periods within our data set.

Furthermore, we use a second dataset to measure the relatedness between industries within our study. The skill-relatedness value is calculated by considering the degree of labour mobility between industries. This data is obtained from an anonymised administrative dataset in Ireland's Central Statistics Office. The dataset contains employee records for each employee within the formal economy. We use this dataset to count the number of workers who transition between industries. Within this dataset industries are classified using the 4-digit NACE 1.1 industry classification. Note, that this is a slightly different industry classification compared to our other dataset. We therefore need to match the two datasets so that industries are classified by the same industry codes. 

To ensure industries are similarly defined we convert the second dataset also into the NACE $2$ industry classification. We do this by following the methodology of \citet{diodato2018network}. We use the correspondence tables between NACE $2$ and NACE $1.1$ released by the European Commission, as well as there detailed documentation of how these classifications differ \citep{correspondencetable}. It is important to note, that there is not a one-to-one correspondence of industries. Hence, as single industry may be classified to various possible other industries (a many-to-many correspondence). We illustrate our conversion process, through the following example: 
\begin{enumerate}
    \item Given that the labour mobility between industry $i$ and $j$ (defined by the NACE $1.1$ industry classification) is $x$.
    \item Then, according to the correspondence table, industry $i$ is now defined as industry $a$ or $b$ (defined by the NACE $2$ industry classification), and industry $j$ as industry $c$. 
    \item We then assume that the labour mobility between industry $a$ and $c$ (defined by the NACE $2$ industry classification), as well as the labour mobility between industry $b$ and $c$ is $x$. 
\end{enumerate}
Note, that our assumption ensures that the skill-relatedness amongst the two new industries pairs has the same relatedness value as the single pair within the original classification.  

\begin{table}[b!]
\caption{Descriptors of the number of industries across time, within regions, and by ownership type within our sample data}
\resizebox{\textwidth}{!}{%
\begin{tabular}{l|ll|ll|ll|ll|ll|ll|ll|ll|ll}
\hline
 & \multicolumn{2}{c|}{All} & \multicolumn{2}{c|}{Border} & \multicolumn{2}{c|}{Dublin} & \multicolumn{2}{c|}{Mid East} & \multicolumn{2}{c|}{Mid West} & \multicolumn{2}{c|}{Midlands} & \multicolumn{2}{c|}{South East} & \multicolumn{2}{l|}{South West} & \multicolumn{2}{c}{West} \\ \hline
Year & DOM & MNE & DOM & MNE & DOM & MNE & DOM & MNE & DOM & MNE & DOM & MNE & DOM & MNE & DOM & MNE & DOM & MNE \\ \hline
2006 & 334 & 180 & 141 & 42 & 221 & 94 & 151 & 55 & 124 & 55 & 79 & 34 & 122 & 40 & 175 & 60 & 153 & 42 \\
2007 & 346 & 181 & 142 & 41 & 221 & 94 & 151 & 52 & 130 & 55 & 82 & 33 & 124 & 38 & 175 & 61 & 157 & 41 \\
2008 & 352 & 177 & 143 & 40 & 226 & 92 & 160 & 51 & 129 & 53 & 83 & 33 & 129 & 38 & 174 & 65 & 157 & 40 \\
2009 & 361 & 176 & 146 & 40 & 230 & 90 & 158 & 52 & 128 & 50 & 85 & 32 & 129 & 38 & 176 & 63 & 156 & 38 \\
2010 & 358 & 175 & 143 & 38 & 228 & 90 & 158 & 51 & 129 & 50 & 90 & 31 & 127 & 36 & 173 & 63 & 146 & 39 \\
2011 & 358 & 177 & 144 & 38 & 225 & 92 & 161 & 50 & 128 & 47 & 89 & 31 & 129 & 37 & 172 & 61 & 140 & 37 \\
2012 & 357 & 178 & 132 & 39 & 225 & 97 & 157 & 51 & 127 & 46 & 84 & 30 & 127 & 38 & 176 & 65 & 141 & 36 \\
2013 & 358 & 180 & 133 & 38 & 228 & 102 & 154 & 50 & 120 & 45 & 84 & 31 & 122 & 35 & 173 & 66 & 141 & 36 \\
2014 & 359 & 186 & 136 & 37 & 227 & 110 & 159 & 50 & 126 & 46 & 85 & 29 & 124 & 36 & 176 & 67 & 145 & 36 \\
2015 & 359 & 190 & 140 & 36 & 232 & 113 & 159 & 47 & 127 & 49 & 88 & 27 & 125 & 38 & 166 & 67 & 144 & 37 \\
2016 & 365 & 196 & 139 & 35 & 234 & 116 & 158 & 48 & 123 & 49 & 85 & 25 & 126 & 40 & 168 & 72 & 148 & 38 \\
2017 & 366 & 197 & 143 & 37 & 233 & 116 & 161 & 49 & 127 & 50 & 86 & 28 & 127 & 40 & 165 & 74 & 147 & 39 \\
2018 & 365 & 202 & 140 & 37 & 233 & 123 & 161 & 52 & 126 & 51 & 86 & 29 & 127 & 40 & 166 & 76 & 150 & 44 \\
2019 & 363 & 203 & 141 & 36 & 230 & 123 & 162 & 52 & 124 & 52 & 85 & 30 & 126 & 43 & 164 & 74 & 149 & 45 \\ \hline
\end{tabular}%
}
\label{Tab_DataDes}
\end{table}

}
\section{Skill-relatedness matrix}\label{SecSR}

In this section, we further elaborate on the methodology used to construct the skill-relatedness measure. This measure calculates the degree of skill and knowledge overlap between industries. 

To quantify the skill-relatedness between industries we follow the recipe of \citet{neffke2017}. Let $F_{ij}$ be the number of job switches between industries $i$ and $j$ (during a given period). As this value is highly dependent on the size of the corresponding industries, it cannot be used as a relatedness measure alone. We therefore compare this value to a baseline: the number of job switches that is expected at random (this corresponds to the configuration null model \citep{molloy1995critical}). This is given by, 
\begin{equation}
   \tilde{SR}(i,j)=\frac{F_{ij}/\sum_j F_{ij}}{\sum_iF_{ij}/\sum_{ij}F_{ij}}.
\end{equation}
The matrix is then normalized and made symmetric by averaging with its transpose and applying a transformation to ensure all values lie between  $-1$ to $1$. Hence,  
\begin{equation}
    A_{SR}(i,j)=\frac{\tilde{SR}(i,j)+\tilde{SR}(j,i)+2}{\tilde{SR}(i,j)+\tilde{SR}(j,i)-2}.
\end{equation}
As we are only interested in cases in which the number of job switches are greater than expected\footnote{This corresponds to a ratio $\tilde{SR}>1$.}, we conserve only positive values of this matrix ($(A_{SR}(i,j)>0)$). 

\section{Further Descriptive Analysis}\label{Appendix_2}

In our study, we investigate how the cohesion to domestic and MNE industries is associated with domestic industry entry and survival. We investigate the cohesion to three mutually exclusive subsets of industries, namely the so-called exclusive MNE industries (industries in which only MNEs are active), the exclusive domestic industries (industries in which only domestic firms are active) and the overlapping industries (those in which both are active). 

In Table~\ref{Tab_Des2} we show the descriptive statistics for both the presence and employment size of domestic industries and MNE industries, as well as those of the three different sets of industries mentioned. We show the descriptive variables for all regions and time periods together. In the table $X$ represents a binary variable indicating whether an industry is present. $Emp$ represents the number of employees within an industry. 

\begin{table}[b]
\caption{Descriptive on industry presence and size for different ownership-type industries}
\centering
\begin{tabular}{l|lllll}
\hline
Variable & N & Mean & SC & Min & Max \\
\hline
$X_D$ & 51632 & 0.3208 & 0.4668 & 0 & 1 \\
$X_M$ & 51632 & 0.1130 & 0.3166 & 0 & 1 \\
$X_{excl D}$ & 51632 & 0.2476 & 0.4316 & 0 & 1 \\
$X_{excl M}$ & 51632 & 0.0398 & 0.1955 & 0 & 1 \\
$X_{overlap}$ & 51632 & 0.0732 & 0.2605 & 0 & 1 \\
$Emp_D$ & 51632 & 42.1126 & 189.3348 & 0 & 8597 \\
$Emp_M$ & 51632 & 47.4123 & 464.8045 & 0 & 23899 \\
$Emp_{excl D}$ & 51632 & 34.0544 & 144.7844 & 0 & 4632 \\
$Emp_{excl M}$ & 51632 & 39.3541 & 418.1570 & 0 & 23008 \\
$Emp_{overlap}$ & 51632 & 16.1163 & 237.0211 & 0 & 17194 \\ 
\hline
\end{tabular}
\label{Tab_Des2}
\end{table}

We observe that the presence of domestic industries is higher than that of MNE industries. We also find that there are, on average, the most exclusive domestic industries present and the least exclusive MNE industries present. What is particularly interesting is that we find that the employment size of multinational industries is larger than that of domestic industries. The set of industries with the largest number of employees on average is found to be the exclusive MNE industries. It is important to remember that this dataset specifically considers a set of industries that are supported by government agencies and not all industries (or employment) within Ireland.

In the paper, we specifically investigate how the presence of MNEs within the same industry, as well as its cohesion to related MNEs impacts domestic firm entry and survival. As a preliminary descriptive step, in Figure~\ref{FigDom2MNE} we illustrate the number of new domestic industries that entered into a exclusive MNE industries within each year $(A)$, across sectors $(B)$ and in different Irish regions $(C)$. We observe an increase in entries particularly in the $2017-18$ time period. We also observe this increase occured across various regions and sectors.

\begin{figure}[t!]
    \centering
    \includegraphics[width = \textwidth]{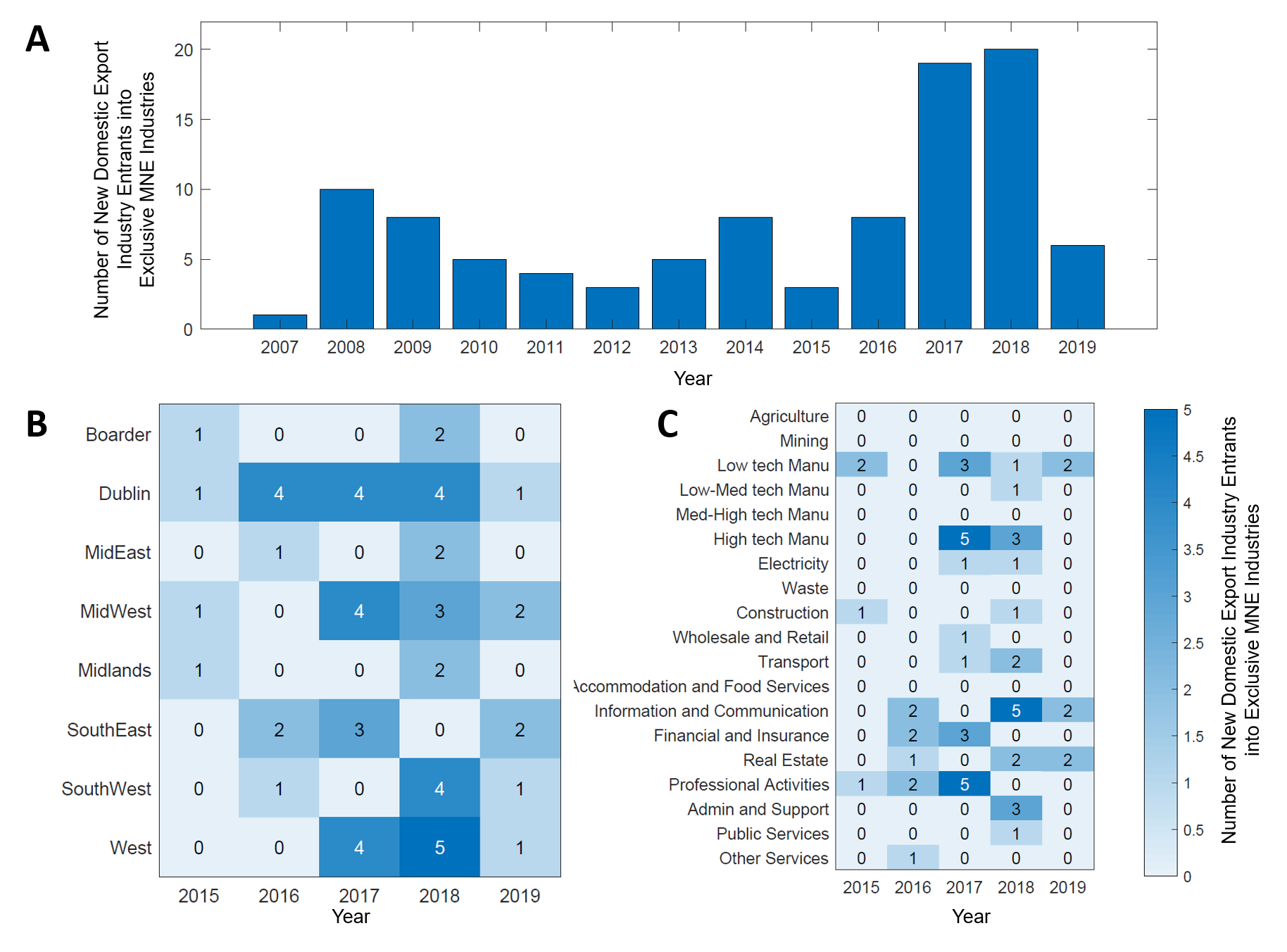}
    \caption{The number of new domestic industry entries into sectors in which only MNEs are active. We show this for $(A)$ different years, $(B)$ different regions and $(C)$ different sectors.}
    \label{FigDom2MNE2}
\end{figure}

\section{Extended Economic Model Results}\label{Appendix_3}

In this section, we show an extension of the results shown and discussed in Section \ref{Sec_Results3} and \ref{Sec_Results4}.

First, as an extension of the results shown in Table~\ref{Tab:IndEntry_RI} and Table~\ref{Tab:IndEntry_SC}, we show how the cohesion, combining both $WC$ and $SC$ together, to domestic and MNE firms is associated with the entry of new domestic firms. Our results are found in Table~\ref{Tab:IndEntry_SC2}. We observe that our results are consistent with those found when considering each cohesion measure separately. Although the values of the coefficient slightly change, they do not loose significance or change sign. As our cohesion measures remain significant when controlling for the corresponding cohesion measure, this demonstrates empirically that our two measures pick up different dimensions of cohesion. 

Similarly, we extend the results shown in Table~\ref{Tab:IndExit_RI}, and Table~\ref{Tab:IndExit_SC}, by investigating how the cohesion (using both the $WC$ and $SC$) to domestic and MNE industries is assocated with the exit of new domestic firms within a region. As before, our cohesion measures remain significant when controlling for the other cohesion measure, and our prior results therefore still hold. 

\begin{landscape}

\begin{table}
\caption{Panel probit regression results for domestic industry entrance between $2006-2019$ and their weighted closeness and strategic closeness measures as independent variable}
\resizebox{1.4\textwidth}{!}{%
\begin{tabular}{l|lllll|lllll|lllll}
\hline
Baseline period & \multicolumn{5}{c|}{2006-2009} & \multicolumn{5}{c|}{2010-2014} & \multicolumn{5}{c|}{2015-2019} \\ \hline
 & \multicolumn{1}{c}{(1)} & \multicolumn{1}{c}{(2)} & \multicolumn{1}{c}{(3)} & \multicolumn{1}{c}{(4)} & \multicolumn{1}{c|}{(5)} & \multicolumn{1}{c}{(1)} & \multicolumn{1}{c}{(2)} & \multicolumn{1}{c}{(3)} & \multicolumn{1}{c}{(4)} & \multicolumn{1}{c|}{(5)} & \multicolumn{1}{c}{(1)} & \multicolumn{1}{c}{(2)} & \multicolumn{1}{c}{(3)} & \multicolumn{1}{c}{(4)} & \multicolumn{1}{c}{(5)} \\ \hline
MNE industry & \begin{tabular}[c]{@{}l@{}}-0.439 \\ (0.315)\end{tabular} & \begin{tabular}[c]{@{}l@{}}-0.507\\ (132.65)\end{tabular} & \begin{tabular}[c]{@{}l@{}}-0.506\\ (0.394)\end{tabular} & \begin{tabular}[c]{@{}l@{}}-0.564\\ (0.335)\end{tabular} & \begin{tabular}[c]{@{}l@{}}-0.777\\ (0.423)\end{tabular} & \begin{tabular}[c]{@{}l@{}}0.021\\ (0.305)\end{tabular} & \begin{tabular}[c]{@{}l@{}}-0.062\\ (0.307)\end{tabular} & \begin{tabular}[c]{@{}l@{}}0.180 \\ (0.327)\end{tabular} & \begin{tabular}[c]{@{}l@{}}0.016\\ (0.306)\end{tabular} & \begin{tabular}[c]{@{}l@{}}0.109 \\ (0.330)\end{tabular} & \begin{tabular}[c]{@{}l@{}}1.673*** \\ (0.451)\end{tabular} & \begin{tabular}[c]{@{}l@{}}1.619***\\ (0.454)\end{tabular} & \begin{tabular}[c]{@{}l@{}}1.706***\\ (0.459)\end{tabular} & \begin{tabular}[c]{@{}l@{}}\textbackslash{}1.583*** \\ (0.477)\end{tabular} & \begin{tabular}[c]{@{}l@{}}1.529***\\ (0.486)\end{tabular} \\
$WC_{excl D}$ &  & \begin{tabular}[c]{@{}l@{}}0.004\\ (0.058)\end{tabular} &  &  & \begin{tabular}[c]{@{}l@{}}-0.010 \\ (0.061)\end{tabular} &  & \begin{tabular}[c]{@{}l@{}}0.072* \\ (0.062)\end{tabular} &  &  & \begin{tabular}[c]{@{}l@{}}0.071*\\ (0.065)\end{tabular} &  & \begin{tabular}[c]{@{}l@{}}0.029\\ (0.118)\end{tabular} &  &  & \begin{tabular}[c]{@{}l@{}}0.064\\ (0.116)\end{tabular} \\
$WC_{excl M}$ &  &  & \begin{tabular}[c]{@{}l@{}}-0.273**\\ (0.185)\end{tabular} &  & \begin{tabular}[c]{@{}l@{}}-0.256* \\ (0.199)\end{tabular} &  &  & \begin{tabular}[c]{@{}l@{}}0.306\\ (0.268)\end{tabular} &  & \begin{tabular}[c]{@{}l@{}}0.355\\ (0.190)\end{tabular} &  &  & \begin{tabular}[c]{@{}l@{}}0.194\\ (0.229)\end{tabular} &  & \begin{tabular}[c]{@{}l@{}}0.148\\ (0.246)\end{tabular} \\
$WC_{overlap}$ &  &  &  & \begin{tabular}[c]{@{}l@{}}0.236***\\ (0.087)\end{tabular} & \begin{tabular}[c]{@{}l@{}}0.241***\\ (0.095)\end{tabular} &  &  &  & \begin{tabular}[c]{@{}l@{}}-0.041\\ (0.082)\end{tabular} & \begin{tabular}[c]{@{}l@{}}0.053\\ (0.098)\end{tabular} &  &  &  & \begin{tabular}[c]{@{}l@{}}0.123\\ (0.117)\end{tabular} & \begin{tabular}[c]{@{}l@{}}0.131\\ (0.124)\end{tabular} \\
$SC_{excl D}$ &  & \begin{tabular}[c]{@{}l@{}}299.09** \\ (132.65)\end{tabular} &  &  & \begin{tabular}[c]{@{}l@{}}374.45**\\ (143.06)\end{tabular} &  & \begin{tabular}[c]{@{}l@{}}359.37*\\ (190.43)\end{tabular} &  &  & \begin{tabular}[c]{@{}l@{}}377.54**\\ (193.07)\end{tabular} &  & \begin{tabular}[c]{@{}l@{}}165.26*\\ (123.18)\end{tabular} &  &  & \begin{tabular}[c]{@{}l@{}}185.51*\\ (120.71)\end{tabular} \\
$SC_{excl M}$ &  &  & \begin{tabular}[c]{@{}l@{}}9.506\\ (74.821)\end{tabular} &  & \begin{tabular}[c]{@{}l@{}}36.301 \\ (77.278)\end{tabular} &  &  & \begin{tabular}[c]{@{}l@{}}-86.686\\ (76.724)\end{tabular} &  & \begin{tabular}[c]{@{}l@{}}-92.201\\ (78.335)\end{tabular} &  &  & \begin{tabular}[c]{@{}l@{}}-22.411\\ (65.699)\end{tabular} &  & \begin{tabular}[c]{@{}l@{}}-191.338\\ (67.028)\end{tabular} \\
$SC_{overlap}$ &  &  &  & \begin{tabular}[c]{@{}l@{}}96.935**\\ (78.581)\end{tabular} & \begin{tabular}[c]{@{}l@{}}25.424*\\ (15.593)\end{tabular} &  &  &  & \begin{tabular}[c]{@{}l@{}}59.343\\ (91.928)\end{tabular} & \begin{tabular}[c]{@{}l@{}}21.475\\ (94.566)\end{tabular} &  &  &  & \begin{tabular}[c]{@{}l@{}}161.53*\\ (136.87)\end{tabular} & \begin{tabular}[c]{@{}l@{}}165* \\ (138.60)\end{tabular} \\
Region FE & Y & Y & Y & Y & Y & Y & Y & Y & Y & Y & Y & Y & Y & Y & Y \\
Industry FE & Y & Y & Y & Y & Y & Y & Y & Y & Y & Y & Y & Y & Y & Y & Y \\
Constant & \begin{tabular}[c]{@{}l@{}}-8.841\\ (8.14e+06)\end{tabular} & \begin{tabular}[c]{@{}l@{}}-37.098\\ (8.14e+06)\end{tabular} & \begin{tabular}[c]{@{}l@{}}-41.753 \\ (8.14e+06)\end{tabular} & \begin{tabular}[c]{@{}l@{}}-20.72 \\ (8.14e+06)\end{tabular} & \begin{tabular}[c]{@{}l@{}}-10.382 \\ (4.45e+06)\end{tabular} & \begin{tabular}[c]{@{}l@{}}-42.689\\ (5.76e+06)\end{tabular} & \begin{tabular}[c]{@{}l@{}}-27.314\\ (5.76e+06)\end{tabular} & \begin{tabular}[c]{@{}l@{}}-30.156 \\ (5.76e+06)\end{tabular} & \begin{tabular}[c]{@{}l@{}}-22.299\\ (5.76e+06)\end{tabular} & \begin{tabular}[c]{@{}l@{}}-33.665 \\ (5.76e+06)\end{tabular} & \begin{tabular}[c]{@{}l@{}}-16.145 \\ (5.76e+06)\end{tabular} & \begin{tabular}[c]{@{}l@{}}-16.853 \\ (5.76e+06)\end{tabular} & \begin{tabular}[c]{@{}l@{}}-17.514 \\ (5.76e+06)\end{tabular} & \begin{tabular}[c]{@{}l@{}}-19.459\\ (5.75e+06)\end{tabular} & \begin{tabular}[c]{@{}l@{}}-15.815\\ (5.76e+06)\end{tabular} \\ \hline
N & 2522 & 2522 & 2522 & 2522 & 2522 & 2494 & 2494 & 2494 & 2494 & 2494 & 2507 & 2507 & 2507 & 2507 & 2507 \\
AUC & 0.9480 & 0.9494 & 0.9481 & 0.9507 & 0.9524 & 0.9700 & 0.9718 & 0.9706 & 0.9701 & 0.9724 & 0.9814 & 0.9817 & 0.9814 & 0.9822 & 0.9828 \\ \hline
\end{tabular}%
}\newline
\small {Notes: Robust standard error in parenthesis; * p<0.1, ** p< 0.05, *p<0.01}
\label{Tab:IndEntry_SC2}
\end{table}


\begin{table}
\caption{Panel probit regression results for domestic industry exits between $2006-2019$ and their weighted closeness and strategic closeness measures as independent variable}
\resizebox{1.4\textwidth}{!}{%
\begin{tabular}{l|lllll|lllll|lllll}
\hline
Baseline period & \multicolumn{5}{c|}{2006-2009} & \multicolumn{5}{c|}{2010-2014} & \multicolumn{5}{c|}{2015-2019} \\ \hline
 & \multicolumn{1}{c}{(1)} & \multicolumn{1}{c}{(2)} & \multicolumn{1}{c}{(3)} & \multicolumn{1}{c}{(4)} & \multicolumn{1}{c|}{(5)} & \multicolumn{1}{c}{(1)} & \multicolumn{1}{c}{(2)} & \multicolumn{1}{c}{(3)} & \multicolumn{1}{c}{(4)} & \multicolumn{1}{c|}{(5)} & \multicolumn{1}{c}{(1)} & \multicolumn{1}{c}{(2)} & \multicolumn{1}{c}{(3)} & \multicolumn{1}{c}{(4)} & \multicolumn{1}{c}{(5)} \\ \hline
MNE industry & \begin{tabular}[c]{@{}l@{}}-0.606**\\ (0.271)\end{tabular} & \begin{tabular}[c]{@{}l@{}}-0.569**\\ (0.295)\end{tabular} & \begin{tabular}[c]{@{}l@{}}-0.622**\\ (0.275)\end{tabular} & \begin{tabular}[c]{@{}l@{}}-0.424**\\ (0.292)\end{tabular} & \begin{tabular}[c]{@{}l@{}}-0.347* \\ (0.324)\end{tabular} & \begin{tabular}[c]{@{}l@{}}-0.140 \\ (0.255)\end{tabular} & \begin{tabular}[c]{@{}l@{}}-0.217\\ (0.262)\end{tabular} & \begin{tabular}[c]{@{}l@{}}-0.182\\ (0.263)\end{tabular} & \begin{tabular}[c]{@{}l@{}}-0.048\\ (0.293)\end{tabular} & \begin{tabular}[c]{@{}l@{}}-0.224 \\ (0.313)\end{tabular} & \begin{tabular}[c]{@{}l@{}}0.025\\ (0.334)\end{tabular} & \begin{tabular}[c]{@{}l@{}}0.004\\ (0.340)\end{tabular} & \begin{tabular}[c]{@{}l@{}}0.094\\ (0.335)\end{tabular} & \begin{tabular}[c]{@{}l@{}}0.108\\ (0.343)\end{tabular} & \begin{tabular}[c]{@{}l@{}}0.118\\ (0.348)\end{tabular} \\
$WC_{excl D}$ &  & \begin{tabular}[c]{@{}l@{}}-0.304***\\ (0.109)\end{tabular} &  &  & \begin{tabular}[c]{@{}l@{}}-0.358*** \\ (0.117)\end{tabular} &  & \begin{tabular}[c]{@{}l@{}}-0.024\\ (0.076)\end{tabular} &  &  & \begin{tabular}[c]{@{}l@{}}-0.083\\ (0.086)\end{tabular} &  & \begin{tabular}[c]{@{}l@{}}-0.100\\ (0.094)\end{tabular} &  &  & \begin{tabular}[c]{@{}l@{}}-0.083 \\ (0.098)\end{tabular} \\
$WC_{excl M}$ &  &  & \begin{tabular}[c]{@{}l@{}}-0.101 \\ (0.205)\end{tabular} &  & \begin{tabular}[c]{@{}l@{}}-0.186\\ (0.219)\end{tabular} &  &  & \begin{tabular}[c]{@{}l@{}}0.317**\\ (0.1769)\end{tabular} &  & \begin{tabular}[c]{@{}l@{}}0.174* \\ (0.0713)\end{tabular} &  &  & \begin{tabular}[c]{@{}l@{}}-0.309\\ (0.183)\end{tabular} &  & \begin{tabular}[c]{@{}l@{}}-0.267\\ (0.191)\end{tabular} \\
$WC_{overlap}$ &  &  &  & \begin{tabular}[c]{@{}l@{}}-0.072*\\ (0.065)\end{tabular} & \begin{tabular}[c]{@{}l@{}}-0.127*\\ (0.117)\end{tabular} &  &  &  & \begin{tabular}[c]{@{}l@{}}-0.375*** \\ (0.105)\end{tabular} & \begin{tabular}[c]{@{}l@{}}-0.367***\\ (0.119)\end{tabular} &  &  &  & \begin{tabular}[c]{@{}l@{}}-0.057\\ (0.079)\end{tabular} & \begin{tabular}[c]{@{}l@{}}-0.085\\ (0.090)\end{tabular} \\
$SC_{excl D}$ &  & \begin{tabular}[c]{@{}l@{}}3.183\\ (284.84)\end{tabular} &  &  & \begin{tabular}[c]{@{}l@{}}271.77\\ (308.26)\end{tabular} &  & \begin{tabular}[c]{@{}l@{}}-296.33*\\ (191.1)\end{tabular} &  &  & \begin{tabular}[c]{@{}l@{}}-252.89* \\ (207.57)\end{tabular} &  & \begin{tabular}[c]{@{}l@{}}-265.74*\\ (218.98)\end{tabular} &  &  & \begin{tabular}[c]{@{}l@{}}-220.02*\\ (203.48)\end{tabular} \\
$SC_{excl M}$ &  &  & \begin{tabular}[c]{@{}l@{}}138.64\\ (100.27)\end{tabular} &  & \begin{tabular}[c]{@{}l@{}}132.76\\ (106.06)\end{tabular} &  &  & \begin{tabular}[c]{@{}l@{}}103.62*\\ (76.318)\end{tabular} &  & \begin{tabular}[c]{@{}l@{}}87.677*\\ (66.138)\end{tabular} &  &  & \begin{tabular}[c]{@{}l@{}}45.455\\ (90.916)\end{tabular} &  & \begin{tabular}[c]{@{}l@{}}29.884 \\ (97.064)\end{tabular} \\
$SC_{overlap}$ &  &  &  & \begin{tabular}[c]{@{}l@{}}-174.62*\\ (108.74)\end{tabular} & \begin{tabular}[c]{@{}l@{}}-169.19*\\ (116.12)\end{tabular} &  &  &  & \begin{tabular}[c]{@{}l@{}}10.241 \\ (100.77)\end{tabular} & \begin{tabular}[c]{@{}l@{}}52.516 \\ (103.55)\end{tabular} &  &  &  & \begin{tabular}[c]{@{}l@{}}-108.78\\ (103.03)\end{tabular} & \begin{tabular}[c]{@{}l@{}}-41.016\\ (111.35)\end{tabular} \\
Region FE & Y & Y & Y & Y & Y & Y & Y & Y & Y & Y & Y & Y & Y & Y & Y \\
Industry FE & Y & Y & Y & Y & Y & Y & Y & Y & Y & Y & Y & Y & Y & Y & Y \\
Constant & \begin{tabular}[c]{@{}l@{}}-1.193**\\ (0.622)\end{tabular} & \begin{tabular}[c]{@{}l@{}}1.152\\ (1.522)\end{tabular} & \begin{tabular}[c]{@{}l@{}}-1.488\\ (1.657)\end{tabular} & \begin{tabular}[c]{@{}l@{}}-0.525\\ (0.730)\end{tabular} & \begin{tabular}[c]{@{}l@{}}0.486\\ (1.641)\end{tabular} & \begin{tabular}[c]{@{}l@{}}-21.305 \\ (3.32e+06)\end{tabular} & \begin{tabular}[c]{@{}l@{}}-13.076\\ (3.32e+06)\end{tabular} & \begin{tabular}[c]{@{}l@{}}-17.537 \\ (3.32e+06)\end{tabular} & \begin{tabular}[c]{@{}l@{}}-10.603 \\ (3.32e+06)\end{tabular} & \begin{tabular}[c]{@{}l@{}}-14.277 \\ (3.32e+06)\end{tabular} & \begin{tabular}[c]{@{}l@{}}-15.163\\ (3.32e+06)\end{tabular} & \begin{tabular}[c]{@{}l@{}}-12.527\\ (3.32e+06)\end{tabular} & \begin{tabular}[c]{@{}l@{}}-15.994 \\ (3.32e+06)\end{tabular} & \begin{tabular}[c]{@{}l@{}}-15.631\\ (3.32e+06)\end{tabular} & \begin{tabular}[c]{@{}l@{}}-12.348 \\ (3.32e+06)\end{tabular} \\ \hline
N & 1166 & 1166 & 1166 & 1166 & 1166 & 1194 & 1194 & 1194 & 1194 & 1194 & 1181 & 1181 & 1181 & 1181 & 1181 \\
AUC & 0.9466 & 0.9512 & 0.9469 & 0.9479 & 0.9539 & 0.9479 & 0.9482 & 0.9521 & 0.9549 & 0.9572 & 0.9650 & 0.9668 & 0.9655 & 0.9657 & 0.9669 \\ \hline
\end{tabular}%
}\newline
\small {Notes: Robust standard error in parenthesis; * p<0.1, ** p< 0.05, *p<0.01}
\label{Tab_IndExits_SC2} 
\end{table}

\end{landscape}

\end{document}